\documentclass[12pt]{article}
\usepackage{enumerate}
\usepackage{caption}
\usepackage{amsmath}	
\usepackage{amssymb}
\usepackage{epic}
\usepackage{epic,eepic}
\usepackage{fancyheadings}
\usepackage[dvips]{graphicx}
\usepackage{graphics}
\usepackage{graphicx}
\usepackage{latexsym}
\usepackage{ulem}
\usepackage{xspace}
\usepackage{lscape}
\usepackage{yhmath}
\usepackage[top=30truemm, bottom=30truemm, left=25truemm, right=25truemm]{geometry}
\allowdisplaybreaks

\usepackage{delarray}

\def\IntKer2{R(K)}

\newtheorem{Remark}{Remark}

\begin{document}


\vspace{5mm}
\centerline{\LARGE{{\bf{Kernel Density Estimation by Genetic Algorithm}}}}
\vspace{5mm}
\begin{center}
Kiheiji NISHIDA
\footnote{General Education Center, Hyogo University of Health Sciences. \\Address: 1-3-6, Minatojima, Chuo-ku, Kobe, Hyogo, 650-8530, JAPAN. E-mail: kiheiji.nishida@gmail.com}
\end{center}
\vskip 10mm
\noindent {\bf{ABSTRACT}}

This study proposes a data condensation method for multivariate kernel density estimation by genetic algorithm. First, our proposed algorithm generates multiple subsamples of a given size with replacement from the original sample. The subsamples and their constituting data points are regarded as {\it{chromosome}} and {\it{gene}}, respectively, in the terminology of genetic algorithm. Second, each pair of subsamples breeds two new subsamples, where each data point faces either {\it{crossover}}, {\it{mutation}}, or {\it{reproduction}} with a certain probability. The dominant subsamples in terms of fitness values are inherited by the next generation. This process is repeated generation by generation and brings the sparse representation of kernel density estimator in its completion. We confirmed from simulation studies that the resulting estimator can perform better than other well-known density estimators.
\vskip 10mm

\section{Introduction}

Suppose that $\mathbf{X}_{i}^{T}=(X_{i1}, X_{i2}, ..., X_{id})$, $i=1,2,...,N$, are $d$-dimensional i.i.d. sample data generated from the true density finction $f(\mathbf{x})$ on $\mathbf{x}^{T} = (x_{1}, x_{2}, ..., x_{d}) \in \mathbb{R}^{d}$. We estimate $f(\mathbf{x})$ by the multivariate kernel density estimator (KDE); its general representation is written as
\begin{eqnarray} \label{def.KDE}
\widehat{f}_{\mathbf{H}_{i}}(\mathbf{x}) = \sum_{i=1}^{N} \alpha_{i} K_{\mathbf{H}_{i}}(\mathbf{x}-\mathbf{X}_{i}),
\end{eqnarray}
where $\mathbf{H}_{i}$, $i = 1,2,...,N$, is a symmetric and positive definite $d$-dimensional bandwidth matrix used for the data $\mathbf{X}_{i}$, $K_{\mathbf{H}_{i}}(\mathbf{x}) = |\mathbf{H}_{i}|^{-\frac{1}{2}}K(\mathbf{H}_{i}^{-\frac{1}{2}} \mathbf{x})$ is a non-negative real valued bounded kernel function, and $\alpha_{i}$, $i = 1,2,...,N$, are the weighting parameters assigned for the data $\mathbf{X}_{i}$. Our aim is to obtain the sparse representation of $\widehat{f}_{\mathbf{H}}(\mathbf{x})$, which involves estimating $f(\mathbf{x})$ by choosing the subset of data points among the given data points $\mathbf{X}_{i}^{T}=(X_{i1}, X_{i2}, ..., X_{id})$, $i=1,2,...,N$, while simultaneously minimizing the estimation error to the greatest extent possible. 

Girolami and He (2003) propose the reduced set density estimator (RSDE), which optimizes the weighting parameters $\alpha_{i}$, $i = 1,2,...,N$, in terms of the integrated squared error (ISE) under the constraint $\alpha_{i} \ge 0$, $i = 1,2,...,N$, $\sum_{i=1}^{N} \alpha_{i} = 1$, assuming the bandwidth matrix is a class of scalar bandwidth matrices $h^{2}\mathbf{I}_{d}$ with its size of $h$ as given. Due to the structure of the optimization problem, its outcome allows $\alpha_{i}=0$ for some $i$'s, realizing the sparse representation of KDE. RSDE is further used for outlier detection and classification in He and Girolami (2004).

In addition to RSDE, Nishida and Naito (2021), associated with Klemel\"a (2007) and Naito and Eguchi (2013), propose the kernel density estimation using stagewise minimization algorithm (SMA) for the purpose. The SMA algorithm first requires a dictionary that consists of kernel functions with scalar bandwidth matrices, where the original i.i.d. sample is randomly split into two disjoint sets; one is used for the means of the kernels and calculating the bandwidths in the dictionary, and the other is used for evaluating the resulting density estimator via an evaluation criterion. Subsequently, the SMA algorithm proceeds in a stagewise manner, choosing a new kernel from the dictionary at each stage to minimize the evaluation criterion of the convex combination of the new kernel and the estimator obtained in the previous stage. Both of these are transformed using an appropriate function before processing the convex combination. Then, the resulting convex combination back-transformed by the function results in the density estimator at the present stage, realizing the sparse representation of KDE. For the evaluation criterion, $U$-divergence is employed, which is similar to the Bregman divergence in Bregman (1967) and Zhang et al. (2009, 2010). This algorithm sorts out the data points used for the density estimator in a stagewise manner improving the evaluation criterion and realizes a sparse representation of KDE when it reaches the final stage.
Simulation studies in Nishida and Naito (2021) confirm that the SMA estimator competes with or sometimes performs better than the RSDE estimator in terms of mean integrated squared error (MISE). Simultaneously, SMA outperforms the RSDE in terms of data condensation ratio (DCR), which refers to the ratio of the number of distinct data points used for the estimator to the original sample size $N$. Both methods exhibit a good performance in terms of data condensation, but the number of data points used in the sparse representation cannot be adjusted by users. In addition, the SMA faces a problem. It updates the estimator stage by stage, taking the choices of words up to the present stage as given, which does not necessarily optimizes its evaluation value among all the possible convex combinations of words.

In this study, we propose a new method using genetic algorithm (GA) for KDE in data condensation. A GA (e.g. Goldberg and Holland 1988; Holland 1989; Forrest 1993; Haupt and Haupt 2004; Sivanandam and Deepa 2008) is a stochastic searching algorithm inspired by natural selection process, such as {\it{inheritance}}, {\it{selection}}, {\it{mutation}}, {\it{crossover}} and {\it{reproduction}}. We find many applications of GA to statistical methods, including Duczmal et al. (2007) for spatial scan statistics, Koukouvinos(2007) for exploring $k$-circulant supersaturated designs, Vovan et al. (2021) for clustering discrete elements, and Brito et al. (2020) for stratified sampling. However, we have not found its application to KDE so far.

We present an outline of our GA for KDE here. Suppose that we have an original sample of size $N$. From this original sample, we make a subsample of size $b$ and repeat this process $B$ times in total with replacement. In our GA, we regard each subsample as vector of data points, where the vector and each of its element are regarded as ${\it{chromosome}}$ and ${\it{gene}}$, respectively, in the terminology of GA. Then, we obtain the initial population of chromosomes of population size $B$. In the subsequent generation, we evaluate each inherited chromosome by calculating ${\it{fitness}}$ of the KDE using the cromsome to the true density and sort the chromosomes in the descending sequence of fitness values. The number of upper $B_{elite}$ chromosomes in the sequence are taken over to the next generation by the {\it{elite selection}} rule. At the same time, we make pairs of chromosomes in such a way that two adjoining chromosomes in the sequence are coupled exhaustively and are mutually exclusive. Then, each pair of chromosomes breed two new chromosomes, where each gene in the chromosome is faced with either crossover, mutation, or reproduction with a certain probability. The new chromosomes are sorted again by the updated fitness values and the number of upper $B-B_{elite}$ of the chromosomes in the sequence are taken over to the next generation. This process is repeated generation by generation until it reaches the final generation $G$. At its completion, we obtain the best fitted subsample and bandwidth matrix for KDE, which have survived all through the selection process. The algorithm is considered to be a solution for combinational optimization problems to extract the subset of data points from the original sample and use the subset to construct the KDE and yield the best fitness value. Generally, a GA is exertive in finding heuristic solutions of a combinational optimization problem.

The remainder of this paper is organized as follows. Section~\ref{alg} states our proposed GA for KDE in data condensation. Section~\ref{Sim} presents simulation studies to validate our proposed GA. Section~\ref{Dis} discusses the results of the study.
%
\section{The proposed GA} \label{alg}

In this section, we describe our proposed GA for KDE. Let $V(\mathbf{D}, \mathbf{H})$ be a fitness function that measures the discrepancy between the true density function $f(\mathbf{x})$ and the KDE in \eqref{def.KDE} which uses the sample $\mathbf{D}=\{ \mathbf{X}_{1}, \mathbf{X}_{2}, ..., \mathbf{X}_{p} \}$ and the universal bandwidth matrix $\mathbf{H} = \mathbf{H}_{1} = \mathbf{H}_{2} = \cdots = \mathbf{H}_{p}$.
\\\\{\bf{Step 1:}} Initial generation $g=1$:
\begin{enumerate}
\item Define the size of subsample $b$, number of subsamples $B$, final generation number $G$, and the weighting parameters $\sum_{i=1}^{b} \beta_{i} = 1$.
\item Make the number of $B$ subsamples of size $b$ with replacement from the original sample of size $N$, where $B$ is an even number. Each subsample is called {\it{chromosome}} and is denoted as $\mathbf{D}_{i}^{(1)} = \{\mathbf{X}_{i, 1}^{(1)}, \mathbf{X}_{i, 2}^{(1)}, ..., \mathbf{X}_{i, b}^{(1)} \}$, $i=1,2,...,B$, where $\mathbf{X}_{i, j}^{(1)}$, $j=1,2,...,b$, is the $j$-th data point of the $i$-th subsample called {\it{gene}}. The {\it{population}} in the generation 1 is written as $\mathbf{D}^{(1)} = \{\mathbf{D}_{1}^{(1)}, \mathbf{D}_{2}^{(1)}, ..., \mathbf{D}_{B}^{(1)} \}$.
\end{enumerate}
{\bf{Step 2:}}
The generations $g=2,3,...,G-1$:
\begin{enumerate}
\item Inherite the population $\mathbf{D}^{(g-1)} = \{\mathbf{D}_{1}^{(g-1)}, \mathbf{D}_{2}^{(g-1)}, ..., \mathbf{D}_{B}^{(g-1)} \}$ from the previous generation $g-1$.
\item For each subsample $\mathbf{D}_{i}^{(g-1)}$, $i=1,2,...,B$, calculate the fitness value $V(\mathbf{D}_{i}^{(g-1)}, \mathbf{H}_{i}^{(g-1)})$ along with the optimal bandwidth matrix $\mathbf{H}_{i}^{(g-1)}$. Then, sort the elements in $\mathbf{D}^{(g-1)} = \{\mathbf{D}_{1}^{(g-1)}, \mathbf{D}_{2}^{(g-1)}, ..., \mathbf{D}_{B}^{(g-1)} \}$ in the descending order of their fitness values $V(\mathbf{D}_{i}^{(g-1)}, \mathbf{H}_{i}^{(g-1)})$, $i=1,2,...,B$, and rename the resulting sequence as $\mathbf{D}^{(g)} = \{\mathbf{D}_{1}^{(g)}, \mathbf{D}_{2}^{(g)}, ..., \mathbf{D}_{B}^{(g)} \}$.
\item Make the replica $\mathbf{D}^{+(g)} \equiv \mathbf{D}^{(g)}$.
\item Breed two new subsamples using the pair of subsamples $\mathbf{D}_{2k-1}^{(g)}$ and $\mathbf{D}_{2k}^{(g)}$, $k=1,2,...,B/2$; each pair of data points $\mathbf{X}_{2k-1, j}^{(g)}$ and $\mathbf{X}_{2k, j}^{(g)}$, $j = 1,2,...,b$, faces either of the following with a certain probability.
\begin{enumerate}[(i)]
\item {\it{Mutation}} : With mutation probability $p_{m}$, $\mathbf{X}_{2k-1, j}^{(g)}$ and $\mathbf{X}_{2k, j}^{(g)}$ are replaced with the two data points randomly chosen from $\mathbf{D}_{1}^{+(g)} = \{\mathbf{X}_{1, 1}^{+(g)}, \mathbf{X}_{1, 2}^{+(g)}, ..., \mathbf{X}_{1, b}^{+(g)} \}$.
\item {\it{Uniform crossover}} : $\mathbf{X}_{2k-1, j}^{(g)}$ is swapped for $\mathbf{X}_{2k, j}^{(g)}$ with crossover probability $p_{u}$.
\item {\it{Reproduction}} : $\mathbf{X}_{2k-1, j}^{(g)}$ and $\mathbf{X}_{2k, j}^{(g)}$ remain unchanged with probability $1-p_{u}-p_{m}$.
\end{enumerate}
\item For each renewed subsample $\mathbf{D}_{i}^{(g)}$, for $i=1,2,...,B$, calculate the fitness value $V(\mathbf{D}_{i}^{(g)}, \mathbf{H}_{i}^{(g)})$ along with the optimal bandwidth matrix $\mathbf{H}_{i}^{(g)}$. Then, sort the renewed subsamples in the descending order of their renewed fitness values, and rename the resulting sequence as $\mathbf{D}^{*(g)} = \{\mathbf{D}_{1}^{*(g)}, \mathbf{D}_{2}^{*(g)}, ..., \mathbf{D}_{B}^{*(g)} \}$.
\item The renewed population $\mathbf{D}^{(g)} = \{\mathbf{D}_{1}^{+(g)}, \mathbf{D}_{2}^{+(g)}, ..., \mathbf{D}_{p_{e}B}^{+(g)}, \mathbf{D}_{1}^{*(g)}, \mathbf{D}_{2}^{*(g)}, ..., \mathbf{D}_{(1-p_{e}) B}^{*(g)} \}$ is taken over to the generation $g+1$, where $p_{e}$ is the ratio of the number of subsamples in $B$ inherited to the next generation by the {\it{elite selection}} rule.
\end{enumerate}
{\bf{Step~3:}} Completion of the algorithm at $g=G$:
\begin{enumerate}
\item Repeat Step~2 stagewise until $g$ reaches $G$. Then, accept the KDE exhibiting the best fitness value $V(\mathbf{D}^{*}, \mathbf{H}^{*})$ written as
\begin{eqnarray}
\widehat{f}_{\mathbf{H}^{*}}(\mathbf{x}) = \sum_{i=1}^{b} \beta_{i} K_{\mathbf{H}^{*}} (\mathbf{x} - \mathbf{X}_{i}^{*}), \nonumber
\end{eqnarray}
along with the resulting subsample $\mathbf{D}^{*} = \{\mathbf{X}_{1}^{*}, \mathbf{X}_{2}^{*}, ..., \mathbf{X}_{b}^{*} \}$ and bandwidth matrix $\mathbf{H}^{*}$.
\end{enumerate}

\begin{figure}[h]
\begin{center}
\includegraphics[width=\textwidth]{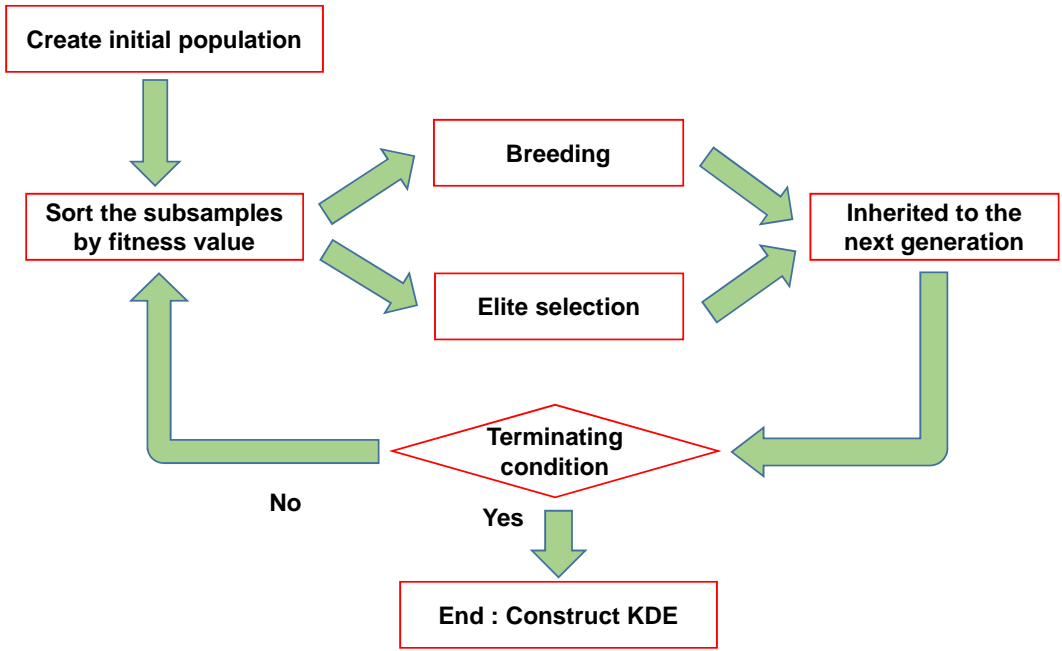}
\caption{The flowchart of our GA.} \label{flowchart}
\end{center}
\end{figure}

We present the flowchart in Figure~\ref{flowchart}.\\\\
{\Remark{\hspace{-2.0mm}{\bf{.}}} \label{remark.sensitivity}} \hspace{-2.0mm} As an empirical rule regarding the choice of $p_{m}$ and $p_{u}$, it is known that setting $p_{m} \le 0.05 \ll p_{u} < 0.5$ is desirable to achieve convergence (Goldberg 1989). If $p_{m}$ is too small, the algorithm tends to yield local minimum solutions; otherwise, it takes a significant amount of time to reach convergence.
{\Remark{\hspace{-2.0mm}{\bf{.}}}} \hspace{-2.0mm} At the phase of mutation in Step~2-4(i), data points can be randomly replaced with the ones in the set of the original sample $\{ \mathbf{X}_{1}, \mathbf{X}_{2}, ..., \mathbf{X}_{N} \}$ instead of the ones in $\mathbf{D}_{1}^{+(g)}$. In such a situation, diversity of the results is retained.
{\Remark{\hspace{-2.0mm}{\bf{.}}} \label{remark.weight}} \hspace{-2.0mm} We define
\begin{eqnarray}
\gamma(i) = \sum_{j=1}^{b} I ( \mathbf{X}_{j}^{*} = \mathbf{X}_{i} ) \beta_{j}, \ \ i=1,2,...,N, \nonumber
\end{eqnarray}
where $I( \cdot )$ is an indicator function that designates unity when the statement in the bracket is true. Then, the resulting KDE is written as
\begin{eqnarray} \label{def.KDE.GA}
\widehat{f}_{\mathbf{H}^{*}}(\mathbf{x}) &=& \sum_{i=1}^{b} \beta_{i} K_{\mathbf{H}^{*}}(\mathbf{x}-\mathbf{X}_{i}^{*}) \nonumber \\
&=& \sum_{i=1}^{N} \gamma(i) K_{\mathbf{H}^{*}}(\mathbf{x}-\mathbf{X}_{i})  \nonumber
\end{eqnarray}
with the data-adaptive weighting parameters $\gamma(i)$, $i=1,2,...,N$, where the algorithm adjusts $\gamma(i)$, $i=1,2,...,N$, to minimize the fitness value, realizing the sparse representation of KDE.
\\\\
{\bf{Fitness function:}} We naturally extend the idea of the least-squares cross-validation method in Rudemo(1982) and Bowman(1984) to our fitness function. Let $\widehat{f}_{\mathbf{H}_{i}}(\mathbf{x}|\mathbf{D}_{i}^{(g)})$ denote KDE using the subsample $\mathbf{D}_{i}^{(g)}$ and the bandwidth $\mathbf{H}_{i}$. We consider ISE between $\widehat{f}_{\mathbf{H}_{i}}(\mathbf{x}| \mathbf{D}_{i}^{(g)})$ and $f(\mathbf{x})$ written as
\begin{eqnarray} \label{def.ISE}
\lefteqn{ISE(\widehat{f}_{\mathbf{H}_{i}}(\cdot|\mathbf{D}_{i}^{(g)}), f(\cdot))} \nonumber \\
&=& \int_{\mathbb{R}^{d}} \biggl[ \widehat{f}_{\mathbf{H}_{i}}(\mathbf{x}|\mathbf{D}_{i}^{(g)}) - f(\mathbf{x}) \biggr]^2 d\mathbf{x} \nonumber \\
&=& \int_{\mathbb{R}^{d}} \widehat{f}_{\mathbf{H}_{i}}^{2}(\mathbf{x}|\mathbf{D}_{i}^{(g)}) d\mathbf{x} - 2 \int_{\mathbb{R}^{d}} \widehat{f}_{\mathbf{H}_{i}}(\mathbf{x}|\mathbf{D}_{i}^{(g)}) f(\mathbf{x}) d\mathbf{x} + \int_{\mathbb{R}^{d}} f^{2}(\mathbf{x}) d\mathbf{x}.
\end{eqnarray}
Replacing the second term in \eqref{def.ISE} by its empirical form, we obtain the fitness function
\begin{eqnarray} \label{def.CV}
-V(\mathbf{D}_{i}^{(g)}, \mathbf{H}_{i}) &=& \int_{\mathbb{R}^{d}} \widehat{f}_{\mathbf{H}_{i}}^{2}(\mathbf{x}|\mathbf{D}_{i}^{(g)}) d\mathbf{x} - \frac{2}{C} {\sum_{k=1}^{N} \sum_{\substack{l=1}}^{b}} I( \mathbf{X}_{k} \neq \mathbf{X}_{l}^{(g)} ) \beta_{l} K_{\mathbf{H}_{i}}(\mathbf{X}_{k} - \mathbf{X}_{l}^{(g)}), \\
C &=& \sum_{k=1}^{N} \sum_{l=1}^{b}  I( \mathbf{X}_{k} \neq \mathbf{X}_{l}^{(g)}), \nonumber
\end{eqnarray}
where the third term in \eqref{def.ISE} is excluded because it is not involved in optimization with respect to $\mathbf{H}_{i}$. We note that the second term in \eqref{def.CV} is the mean of $\widehat{f}_{\mathbf{H}_{i}}(\mathbf{x}|\mathbf{D}_{i}^{(g)})$ over {\it{test data}} $\mathbf{X}_{1}, \mathbf{X}_{2}, ..., \mathbf{X}_{N}$ while {\it{training data}} $\mathbf{X}_{1}^{(g)}, \mathbf{X}_{2}^{(g)}, ..., \mathbf{X}_{b}^{(g)}$ are left out. If we employ the training data for the averaging in the second term of \eqref{def.CV}, the resulting fitness value is degenerated because the distribution of the training data through our GA is not necessarily identical to that of the original data to be presented in simulation sections. We employ (\ref{def.CV}) as the fitness function to evaluate the estimator and use it to find the optimal bandwidth matrix.
\section{Simulation studies} \label{Sim}

We conduct {{Simulations~1-3}} to validate performance of our proposed GA. Simulations~1 and 2 are Monte-Carlo simulations dealing with the bivariate and trivariate settings respectively. Simulation~3 is a real data application of our method. We assume the bandwidth matrix in the simulations as a class of scalar matrices $h^{2} \mathbf{I}_{d}$ and the weighting parameters are set to be $\beta_{i} = 1/b, i=1,2,...,b$. The competitors to our GA are the Direct Plug-in method in Duong and Hazelton (2003) in the setting of full bandwidth matrix (henceforce, DPI) and the RSDE. In calculating DPI, we employ {\texttt{Hpi}} function in '{\texttt{ks}}' library in R. For kernel, we employ Gaussian throughout the study.

In simulations~1 and 2, we generate a dataset of size $N$ from $f(\mathbf{x})$ and obtain $t=10$ different kernel density estimators by applying our GA $t=10$ times for the dataset. Then, we define the performance measure to be the average of the $t=10$ different ISE's computed by each estimator; we denote the measure to be ISE$^{*}$ or ISE$^{*}(g)$, the latter of which represents the ISE$^{*}$ at the generation $g$. Introducing ISE$^{*}$ for performance measure is inevitable because GA is a stochastic search algorithm. Each simulation is categorized by the different sizes of $b$ because we expect the performance of our GA to be concerned with the size of $b$. In evaluating the performance of DPI and RSDE, we generate the number of $s=10$ datasets of size $N$ and compute MISE by averaging ISE's computed by each dataset, where one of the $s=10$ datasets is identical to the one used for computing the ISE$^{*}$. The simulations are developed in terms of the degree of data condensation as well because we expect the influential data points to the estimation error to survive the selection process of our GA. 
\subsection{Simulation~1 (bivariate)} \label{Sim1}

\begin{figure}[h]
\begin{center}
    \begin{minipage}[t]{0.3\hsize}
        \center
        \captionsetup{width=.95\linewidth}
        \includegraphics[width=\textwidth]{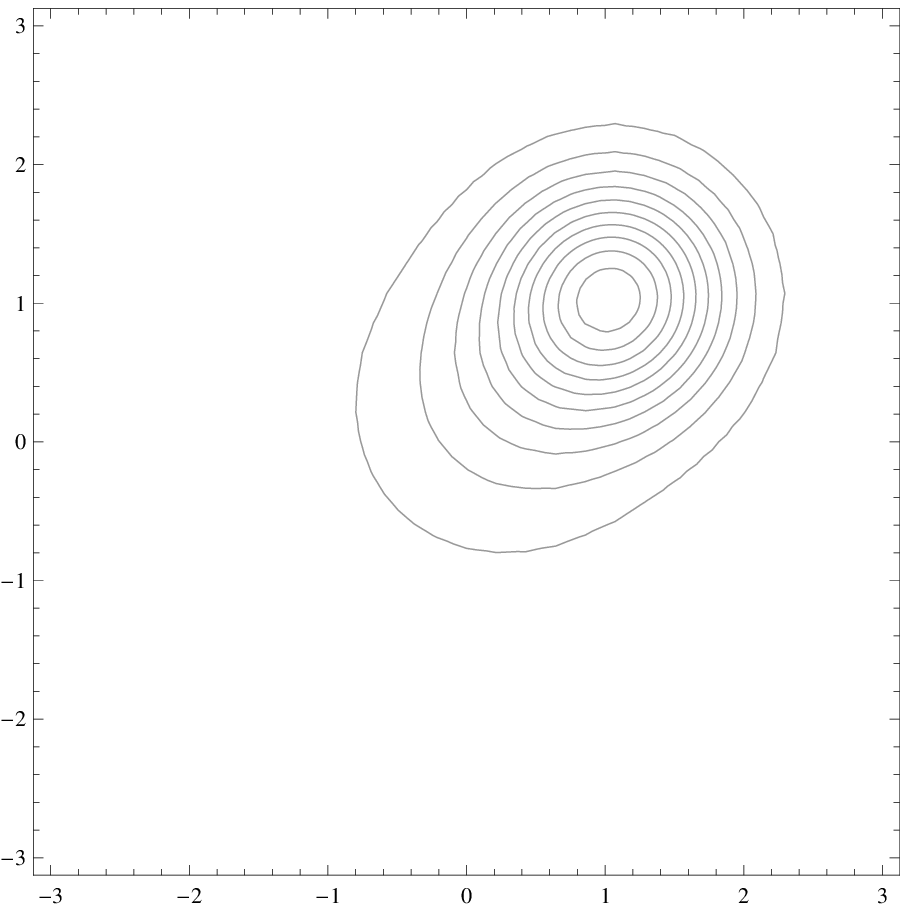}
        \mbox{Type C}
    \end{minipage}
    \begin{minipage}[t]{0.3\hsize}
        \center
        \captionsetup{width=.95\linewidth}
        \includegraphics[width=\textwidth]{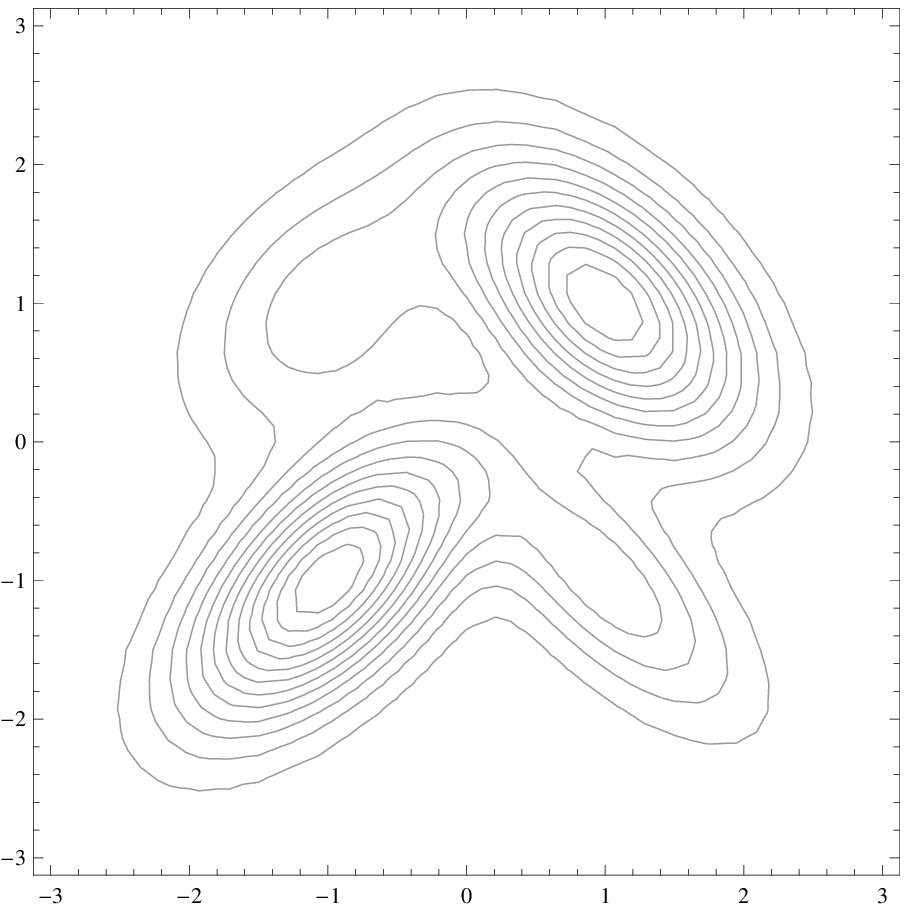}
        \mbox{Type L}
    \end{minipage}
\caption{Simulation~1: True denisty functions (Wand and Jones 1993).} \label{true.C.J.L}
\end{center}
\end{figure}

In Simulation~1, we employ the bivariate simulation settings Type~C and L in Wand and Jones (1993) given below, where the notation $N(\mu_{1}, \mu_{2}, \sigma_{1}^{2}, \sigma_{2}^{2}, \rho)$ represents the bivariate normal density with means $\mu_{1}$, $\mu_{2}$, variances $\sigma_{1}^{2}$, $\sigma_{2}^{2}$, and correlation coefficient $\rho$, whose contour-plots are drawn in Figure~\ref{true.C.J.L}.
\\\\{\bf{Type C. Skewed:}}\\
$\frac{1}{5} N\bigl(0,0,1,1,0\bigr) + \frac{1}{5} N\bigl(\frac{1}{2}, \frac{1}{2}, \bigl(\frac{2}{3}\bigr)^{2}, \bigl(\frac{2}{3}\bigr)^{2}, 0 \bigr) + \frac{3}{5} N \bigl(\frac{13}{12}, \frac{13}{12}, \bigl(\frac{5}{9}\bigr)^{2}, \bigl(\frac{5}{9}\bigr)^{2}, 0 \bigr)$\\\\
{\bf{Type L. Quadrimodal:}}\\
$\frac{1}{8} N \bigl(-1, 1, \bigl(\frac{2}{3}\bigr)^{2}, \bigl(\frac{2}{3}\bigr)^{2}, \frac{2}{5} \bigr) + \frac{3}{8} N \bigl(-1, 1, \bigl(\frac{2}{3}\bigr)^{2}, \bigl(\frac{2}{3}\bigr)^{2}, \frac{3}{5} \bigr) + \frac{1}{8} N \bigl(-1, 1, \bigl(\frac{2}{3}\bigr)^{2}, \bigl(\frac{2}{3}\bigr)^{2}, -\frac{7}{10} \bigr) + \frac{3}{8} N \bigl(1, 1, \bigl(\frac{2}{3}\bigr)^{2}, \bigl(\frac{2}{3}\bigr)^{2}, -\frac{1}{2} \bigr)$\\\\

We employ the following GA parameter settings $(B, G, p_{u}, p_{m}, p_{e})=(50, 100, 0.475, 0.05, 0.1)$ and consider three sample sizes, $N=200, 400$, and $1000$. The numerical results in terms of estimation error for Types~C and L are presented in Tables~\ref{C.results} and \ref{L.results} respectively. The numbers DPI$^{*}$ and RSDE$^{*}$ are the values of ISE $\times 10^{5}$ calculated by the DPI and the RSDE respectively using the identical sample in calculating ISE$^{*}$ by our GA. The minimum values of ISE$^{*}(g)$ at each generation $g$ over the sizes of $b$ are underlined in the Tables. All the parenthetic numbers in the tables represent standard deviations (S.D.) throughout the study. \\\\
{\bf{Summary of the results : Type C}}

\begin{figure}[]
\begin{center}
    \begin{minipage}[t]{0.49\hsize}
        \center
        \captionsetup{width=.95\linewidth}
        \includegraphics[width=\textwidth]{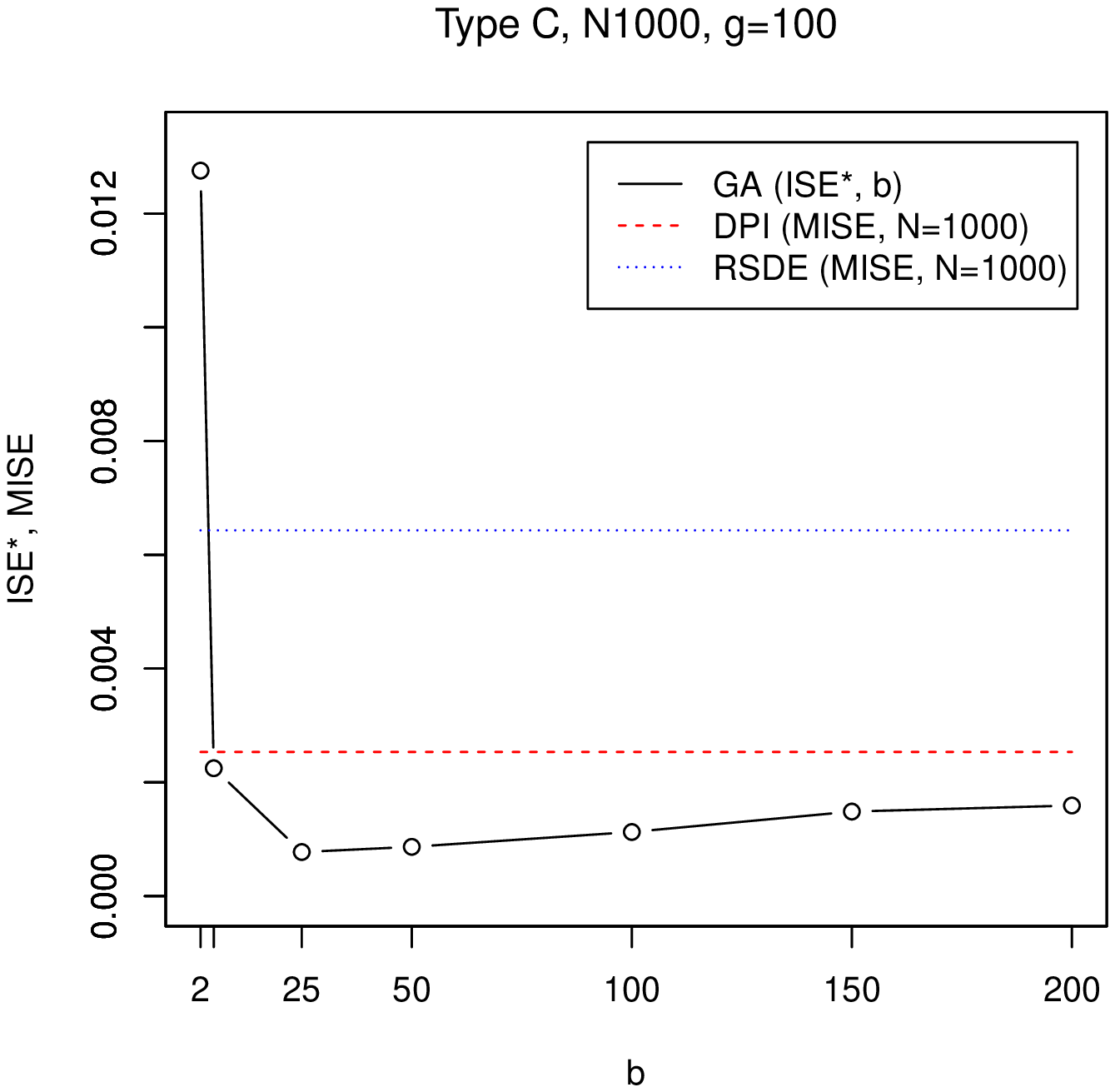}
        \mbox{(a)}
    \end{minipage}
    \begin{minipage}[t]{0.49\hsize}
        \center
        \captionsetup{width=.95\linewidth}
        \includegraphics[width=\textwidth]{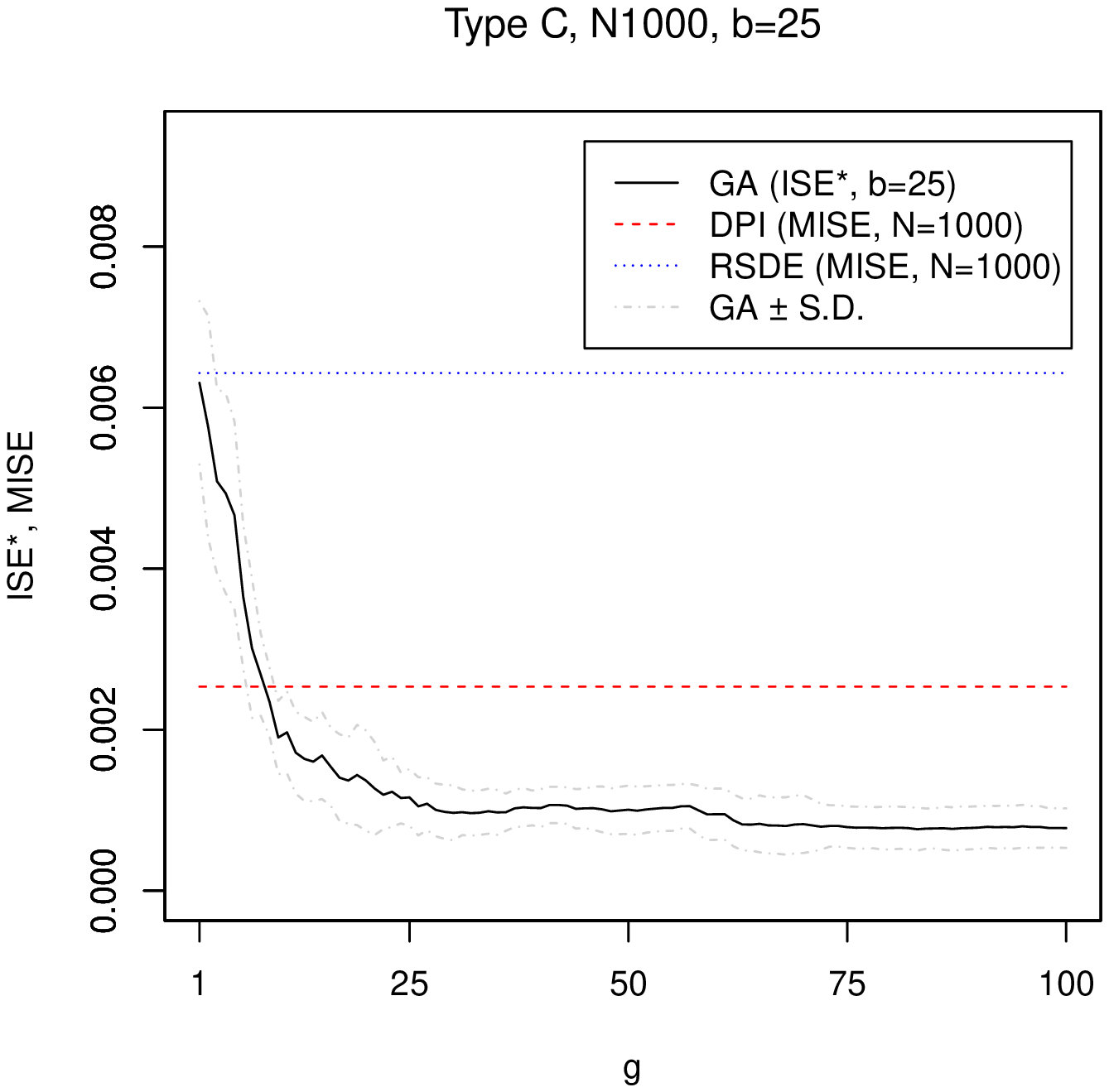}
        \mbox{(b)}
    \end{minipage}\\
    \begin{minipage}[t]{0.49\hsize}
        \center
        \captionsetup{width=.95\linewidth}
        \includegraphics[width=\textwidth]{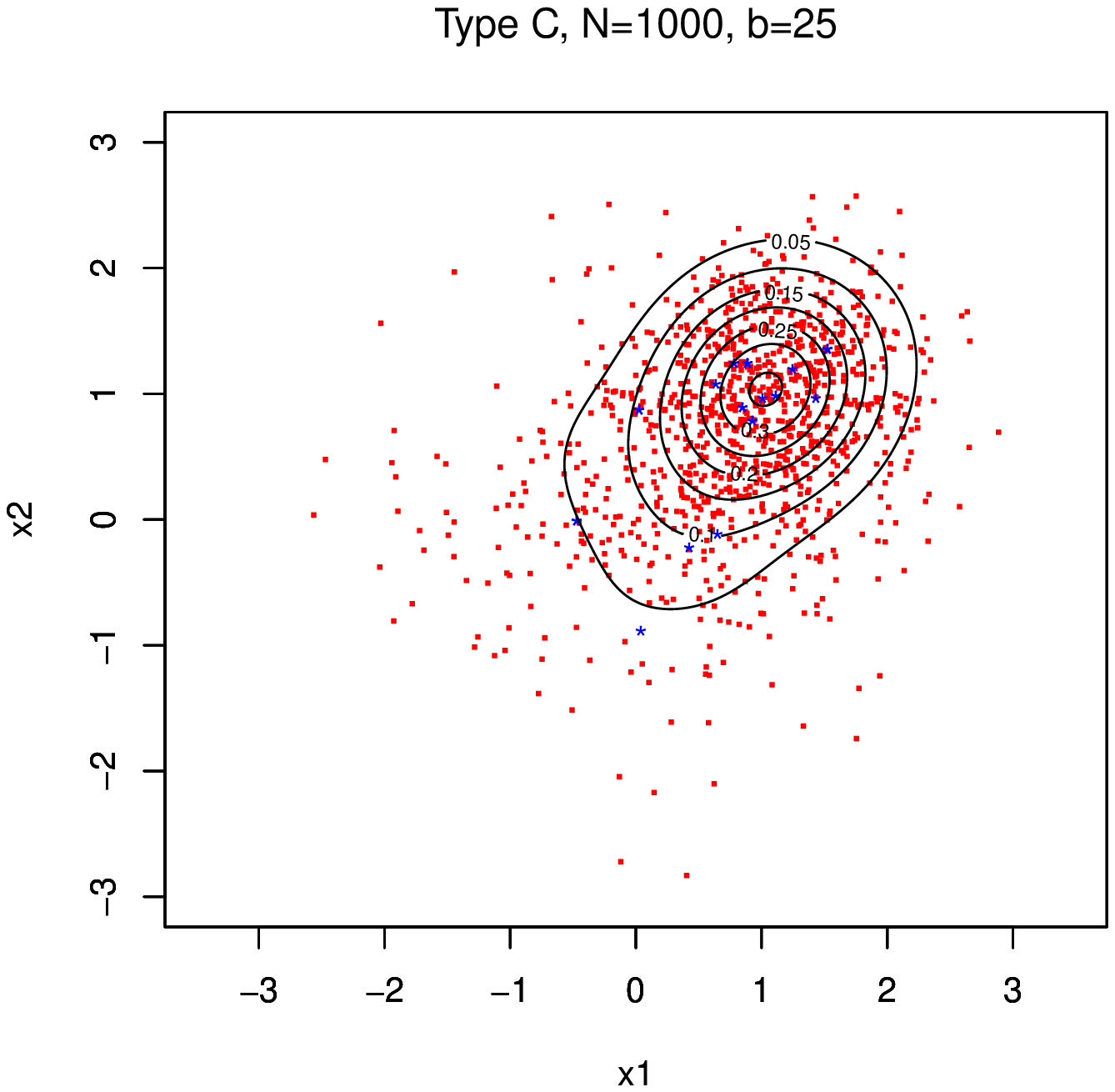}
        \mbox{(c)}
    \end{minipage}
    \begin{minipage}[t]{0.49\hsize}
        \center
        \captionsetup{width=.95\linewidth}
        \includegraphics[width=\textwidth]{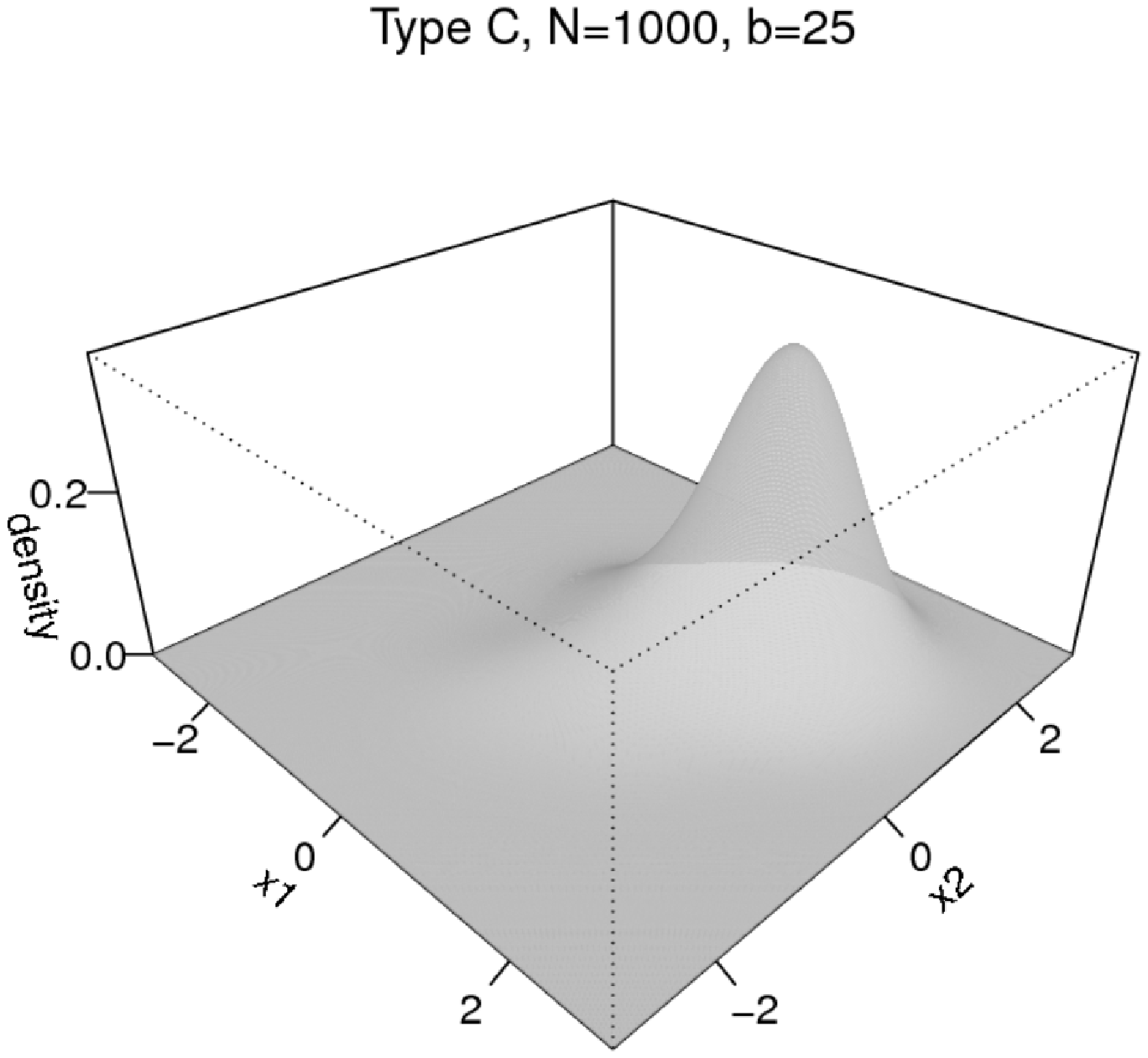}
        \mbox{(d)}
    \end{minipage}
\end{center}
\caption{[Simulation~1 (Type C, $N=1000$)] (a) : Estimation error vs $b$ at $g=100$. (b) : Estimation error vs $g$ at $b=25$. (c) : Contour-plot with the original data points (red) and the data points choosen by the GA methods (blue). (d) : Perspective-plot.} \label{sim1.N1000.C}
\end{figure}

Figure~\ref{sim1.N1000.C} summarizes the results in Table~\ref{C.results} for the case of $N=1000$, where we pick up the case as one example because we observe the common tendencies regardless of $N$. The panel (a) in Figure~\ref{sim1.N1000.C} visually summarizes the results in Table~\ref{C.results} with respect to the size of $b$, representing the plot of the value of ISE$^{*}$ at $g=100$ for each size of $b$, along with the values of MISE for the DPI and the RSDE estimators. We observe that ISE$^{*}(g)$ takes the minimum value in the neighborhood of $b = 25$. It is because the two effects are balanced when $b$ is close to 25: one effect is that ISE$^{*}$ is improved as the number of training data points $b$ increases and the other is that ISE$^{*}$ degenerates as the number of test data points $N-b$ decreases. We also note that our GA can outperform the DPI estimator even in the case of $b=5$ when $g$ is greater than the vicinity of $25$. We consider the five data points chosen by our GA are essential to constructing the density estimator.

The panel (b) in Figure~\ref{sim1.N1000.C} summarizes the result in Table~\ref{C.results} with respect to $g$ when $b=25$. We observe that our proposed GA of $b=25$ can outperform the DPI estimator using the original sample of the size $N=1000$ in terms of estimation error when $g$ is greater than the vicinity of 5. This feature is observed regardless of $N$, excepting the case of $b=2$.

The panels (c) and (d) in Figure~\ref{sim1.N1000.C} are respectively the contour-plot and the perspective-plot of our estimator, both of which comes from one shot of 10 implementations of our GA to a sample, where the red points in the contour-plot are the original sample of the size $N$, while the blue ones are those chosen by our GA in the estimation. We observe that our GA captures the shape of the true density. We also observe that our GA tends to choose the data points close to the peak of the true density, comparing the distribution of the blue points with that of the red ones in the panel (c).

Looking at the values of the cases $(N, b) = (200, 100), (400, 100)$, $(400, 150)$ and $(1000, 150)$ in Table~\ref{C.results}, we notice that ISE$^{*}(g)$ is minimized in the vicinity of $g=25$ over $g$. We consider it to be so because the data points without good influence can be included in the subsample as $g$ progresses when the variety of the training data points is rich.

The numerical results in terms of data condensation for Type~C are presented in Table~\ref{tab.c.ratio.C}. The numbers in column (I) represent the number of distinct data points used for estimating the density on the average over $t=10$ implementations of our GA to a sample. The numbers in column (II) represent the maximum multiplicity of the data points on the average, where the multiplicity is defined to be the number of times the single data point is chosen by the GA. The DCR by our GA is defined to be (I) divided by $N$ on the average over $t=10$ implementations of our GA to a sample. In contrast, the DCR by RSDE is the average of DCR over $s=10$ different datasets, in which the identical dataset used in calculating the DCR by our GA is included.

We observe that our estimator can outperform RSDE in terms of the DCR and yields the smaller DCR as the size $b$ becomes smaller for each sample size $N$. We also observe that the value of (II)$/b$, which means the maximum weight of data points used for the estimation, increases as the size of $b$ becomes small. It means that our GA chooses the data points with great influence on estimation, many times as $b$ becomes small. If we compare the DCR over different sample sizes for each size of $b$, we observe the smaller DCR for the larger sample size $N$.

We are also interested in the performance of the estimator by our GA over different original datasets. For reference, we calculate the MISE of our proposed estimator using the identical original datasets for calculating the MISE's of the DPI and the RSDE in Table~\ref{C.results}. We calculate MISE in such a way that we implement our GA once for each of 10 datasets until $g=100$ and calculate the average of all the resulting ISEs. In the case of $(N, b) = (1000, 25)$, the best performance case in terms of ISE$^{*}$ of the simulation, MISE$ \times 10^{5}$ (S.D. $ \times 10^{5}$) comes out to be $120$ (44), outperforming its competitors. The corresponding DCR (S.D.) is 0.0154 (0.0034). The superiority of our GA over different original samples to its competitors is widely observed in the rest of the cases as well.
\begin{table}
\begin{center}
{\scriptsize{
\begin{tabular}{llllllllll}
\hline
\hline
$g$ & 1 & 25 & 50 & 75 & 100 & DPI &  RSDE & DPI$^{*}$ & RSDE$^{*}$ \\
\hline
$\underline{N=200}$ & --- & --- & --- & --- & --- &  711\ (198) & 1503\ (616) & 578 & 972 \\
$b=2$ & 1855\ (253) & 1485\ (190) & 1485\ (190) & 1485\ (190) & 1485\ (190) \\

$b=5$ & 1014\ (190) & 296\ \ \ (84) & 295\ \ \ (83) & 295\ \ \ (83) & 295\ \ \ (83) \\
$b=25$ & 736\ \ \ (295) & \underline{211}\ \ \ (36) & \underline{197}\ \ \ (31) & \underline{186}\ \ \ (27) & \underline{198}\ \ \ (33) \\
$b=50$ & 568\ \ \ (121) & 223\ \ \ (40) & 215\ \ \ (57) & 210\ \ \ (59) & 215\ \ \ (37) \\
$b=100$ & 587\ \ \ (136) & 266\ \ \ (104) & 270\ \ \ (62) & 289\ \ \ (61) & 277\ \ \ (45) \\
$b=150$ & \underline{448}\ \ \ (114) & 268\ \ \ (98) & 291\ \ \ (64) & 294\ \ \ (98) & 361\ \ \ (95) \\
\hline
$\underline{N=400}$ & --- & --- & --- & --- & --- &  435\ (83) & 1231\ (347) & 413 & 842 \\
$b=2$ & 1523 (375) & 1173 (143) & 1173 (143) & 1173 (143) & 1173 (143) \\

$b=5$ & 894\ \ \ (362) & 347\ \ \ (59) & 327\ \ \ (63) & 327\ \ \ (63) & 327\ \ \ (63) \\
$b=25$ & 742\ \ \ (249) & \underline{309}\ \ \ (38) & \underline{309}\ \ \ (67) & \underline{300}\ \ \ (31) & \underline{293}\ \ \ (31) \\
$b=50$ & 606\ \ \ (173) & 316\ \ \ (58) & 312\ \ \ (51) & 325\ \ \ (46) & 312\ \ \ (31) \\
$b=100$ & \underline{427}\ \ \ (209) & 322\ \ \ (37) & 359\ \ \ (61) & 327\ \ \ (39) & 352\ \ \ (43) \\
$b=150$ & 473\ \ \ (147) & 319\ \ \ (48) & 360\ \ \ (47) & 372\ \ \ (68) & 385\ \ \ (58) \\
\hline
$\underline{N=1000}$ & --- & --- & --- & --- & --- &  253\ (52) & 643\ (190) & 220 & 842 \\
$b=2$ & 1576\ (286) & 1276\ (64) & 1275\ (64) & 1275\ (64) & 1275\ (64) \\

$b=5$ & 860 \ \ (304) & 237 \ \ (63) & 224 \  \ (56) & 224 \ \ (56) & 224 \ \ (56) \\
$b=25$ & 630 \ \ (101) & 112 \ \ (35) & 99 \  \ \ (30) & \underline{78} \ \ \ (25) & \underline{77} \ \ \ (24) \\
$b=50$ & 466 \ \ (140) & \underline{109} \ \ (39) & \underline{79} \  \ \ (24) & 82 \ \ \ (29) & 86 \ \ \ (23) \\
$b=100$ & 373 \ \ (123) & 114 \ \ (28) & 109 \ \ (16) & 115 \ \ (30) & 112 \ \ (22) \\
$b=150$ & \underline{368} \ \ (62) & 133 \ \ (32) & 137 \ \ (39) & 138 \ \ (27) & 148 \ \ (42) \\
\hline
\hline
\end{tabular}}}
\caption[]
{[Simulation~1 (Type C)] Results of estimation error ISE$^{*}(g)$ $\times 10^{5}$ (S.D. $\times 10^{5}$). The numbers in the columns of DPI and RSDE are MISE $\times 10^{5}$ (S.D. $\times 10^{5}$). The numbers in the columns of DPI$^{*}$ and RSDE$^{*}$ are ISE $\times 10^{5}$ of each estimator calculated by the identical sample used in calculating ISE$^{*}$. The minimum values of ISE$^{*}(g)$ at each $g$ over the sizes of $b$ are underlined. } \label{C.results}
\end{center}
\end{table}
\begin{table}
\begin{center}
{\scriptsize{
\begin{tabular}{l|llll|l}
\hline
\hline
& (I) & (II) & (II)$/b\ \ \ $ & DCR.GA & DCR.RSDE \\
\hline
$\underline{N=200}$ & ---  & ---  & --- & --- & .2650\ (.0291) \\
$b=25$ & 11.80 (1.75) & \ \ 5.30 (0.94) & 0.21 & .0590 (.0088) & --- \\
$b=50$ & 19.40 (1.50) & \ \ 7.40 (1.26) & 0.14 & .0970 (.0075) & --- \\
$b=100$ & 32.60 (2.98)  & 12.20 (4.02)  & 0.12 & .1630 (.0149) & ---  \\
$b=150$ & 42.90 (4.79) & 13.80 (1.81) & 0.09 & .2145 (.0240) & ---  \\
\hline
$\underline{N=400}$ & ---  & ---  & --- & --- & .2047\ (.0158) \\
$b=25$ & 14.40 (2.27)  & \ \ 5.30 (2.11) & 0.21 & .0360 (.0057) & ---  \\
$b=50$ & 26.80 (4.61)  & \ \ 6.20 (2.34)  & 0.12 & .0670 (.0115) & ---  \\
$b=100$ & 44.10 (4.58)  & \ \ 7.80 (2.14)  & 0.07 & .1103 (.0115) & ---  \\
$b=150$ & 56.30 (4.85)  & 10.40 (2.87)  & 0.06 & .1408 (.0121) & ---  \\
\hline
$\underline{N=1000}$ & ---  & ---  & --- & --- & .1524\ (.0155) \\
$b=25$ & 15.80 (1.54)  & \ \ 3.60 (0.69)  & 0.14 & .0158 (.0015) & ---   \\
$b=50$ & 29.60 (5.03)  & \ \ 5.40 (1.83)  & 0.10 & .0296 (.0050) & ---   \\
$b=100$ & 56.40 (5.27) & \ \ 5.70 (1.05) & 0.05 & .0564 (.0053) & ---   \\
$b=150$ & 78.30 (8.94) & \ \ 7.60 (1.17) & 0.05 & .0783 (.0089) & ---   \\
\hline
\hline
\end{tabular}}}
\caption[]
{[Simulation~1 (Type C)] (I) : The number of distinct data points used to estimate the density on the average (S.D.). (II) : The maximum multiplicity of data points on the average (S.D.).} \label{tab.c.ratio.C}
\end{center}
\end{table}
\clearpage
\hspace{-8mm} {\bf{Summary of the results : Type L}}

\begin{figure}[]
\begin{center}
    \begin{minipage}[t]{0.49\hsize}
        \center
        \captionsetup{width=.95\linewidth}
        \includegraphics[width=\textwidth]{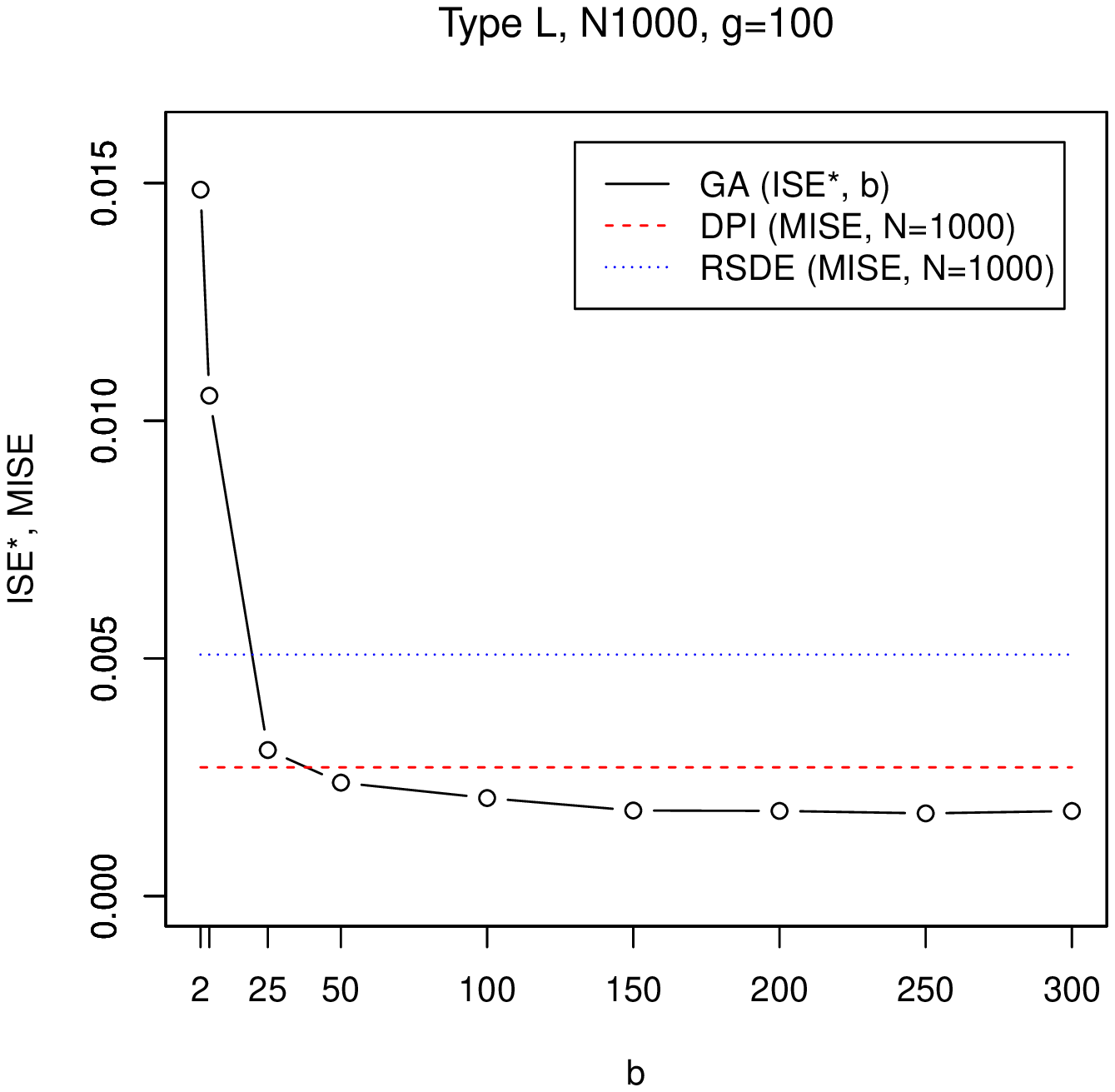}
        \mbox{(a)}
    \end{minipage}
    \begin{minipage}[t]{0.49\hsize}
        \center
        \captionsetup{width=.95\linewidth}
        \includegraphics[width=\textwidth]{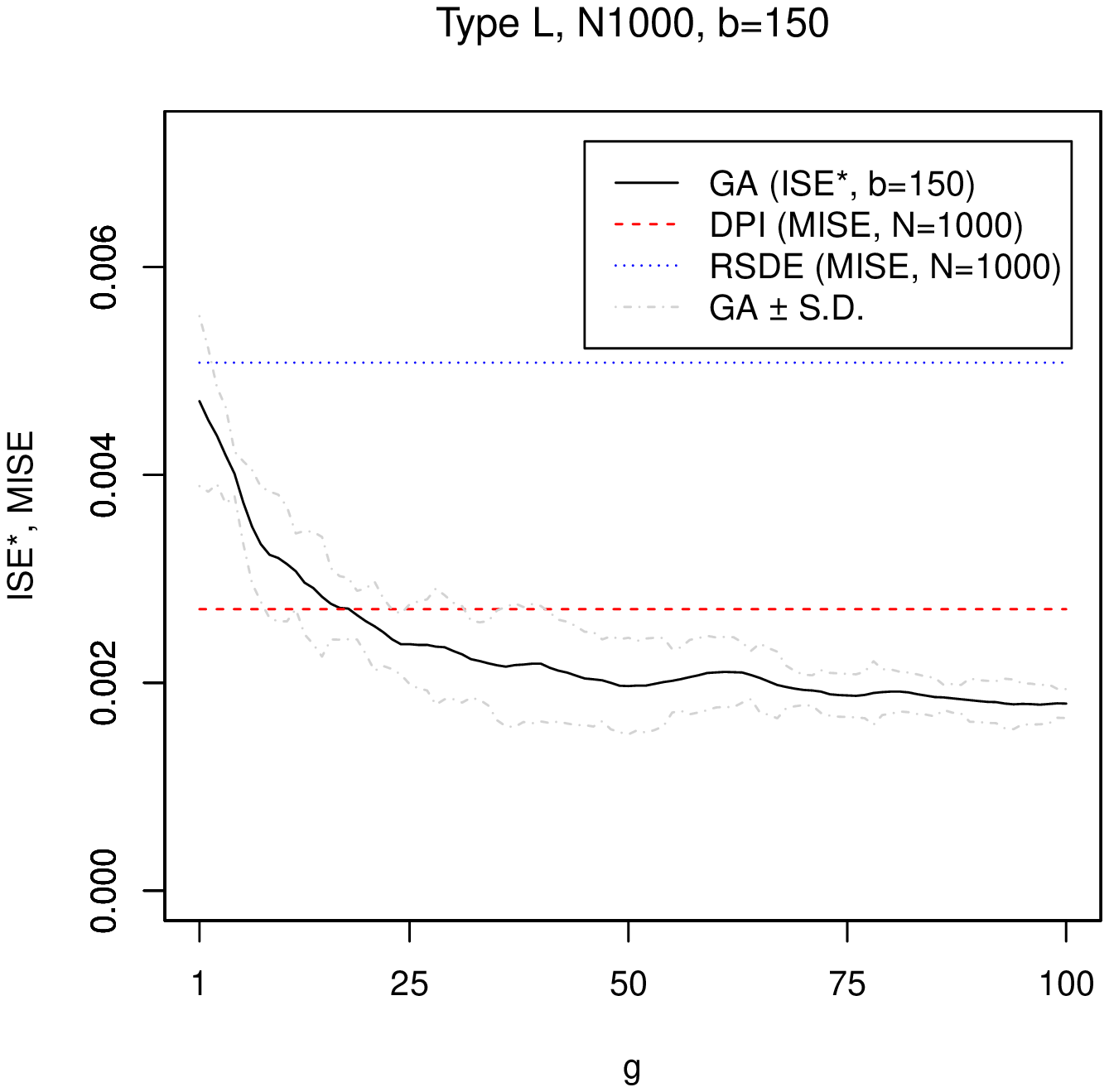}
        \mbox{(b)}
    \end{minipage}\\
    \begin{minipage}[t]{0.49\hsize}
       \center
        \captionsetup{width=.95\linewidth}
        \includegraphics[width=\textwidth]{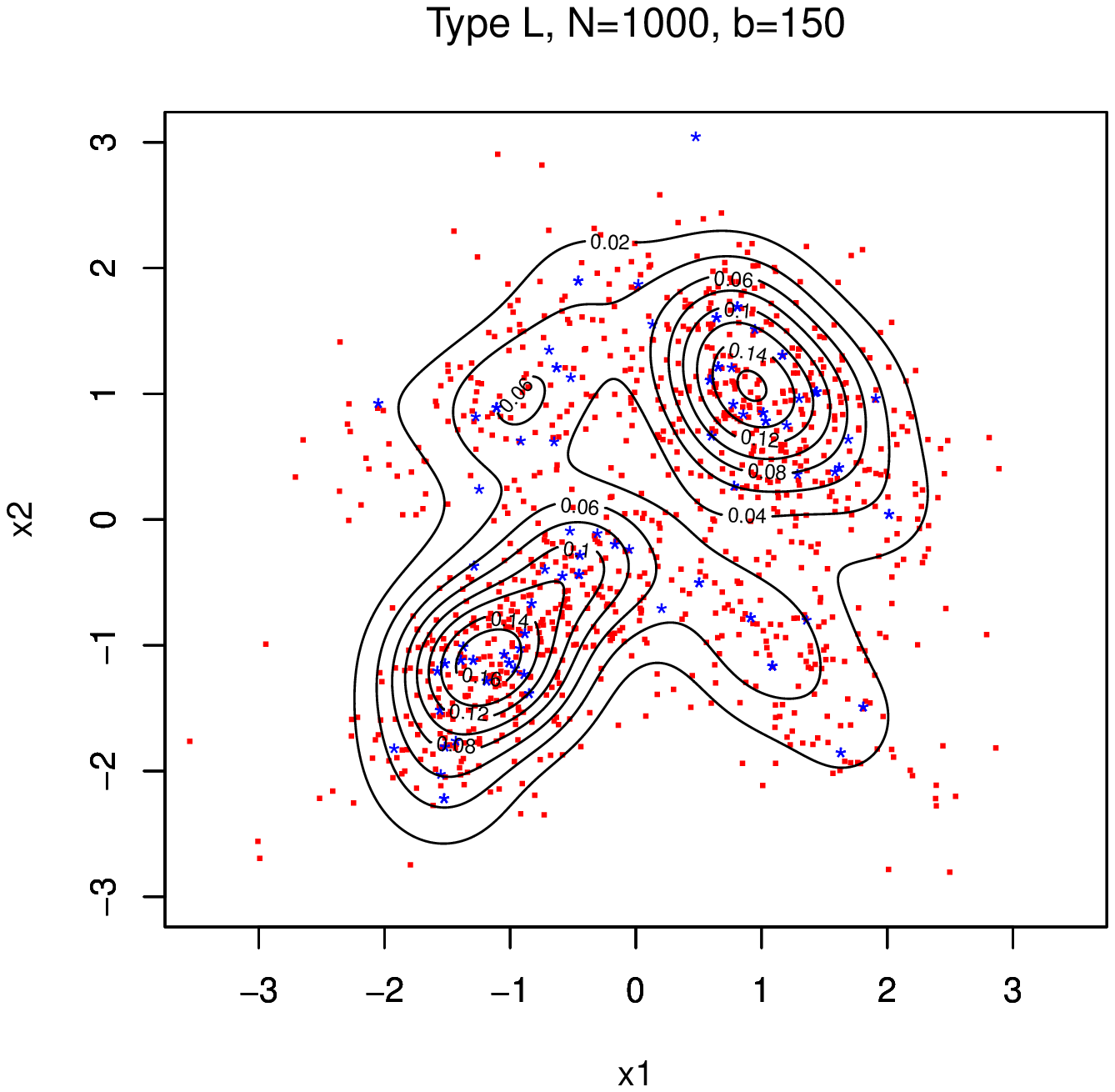}
        \mbox{(c)}
    \end{minipage}
    \begin{minipage}[t]{0.49\hsize}
        \center
        \captionsetup{width=.95\linewidth}
        \includegraphics[width=\textwidth, clip]{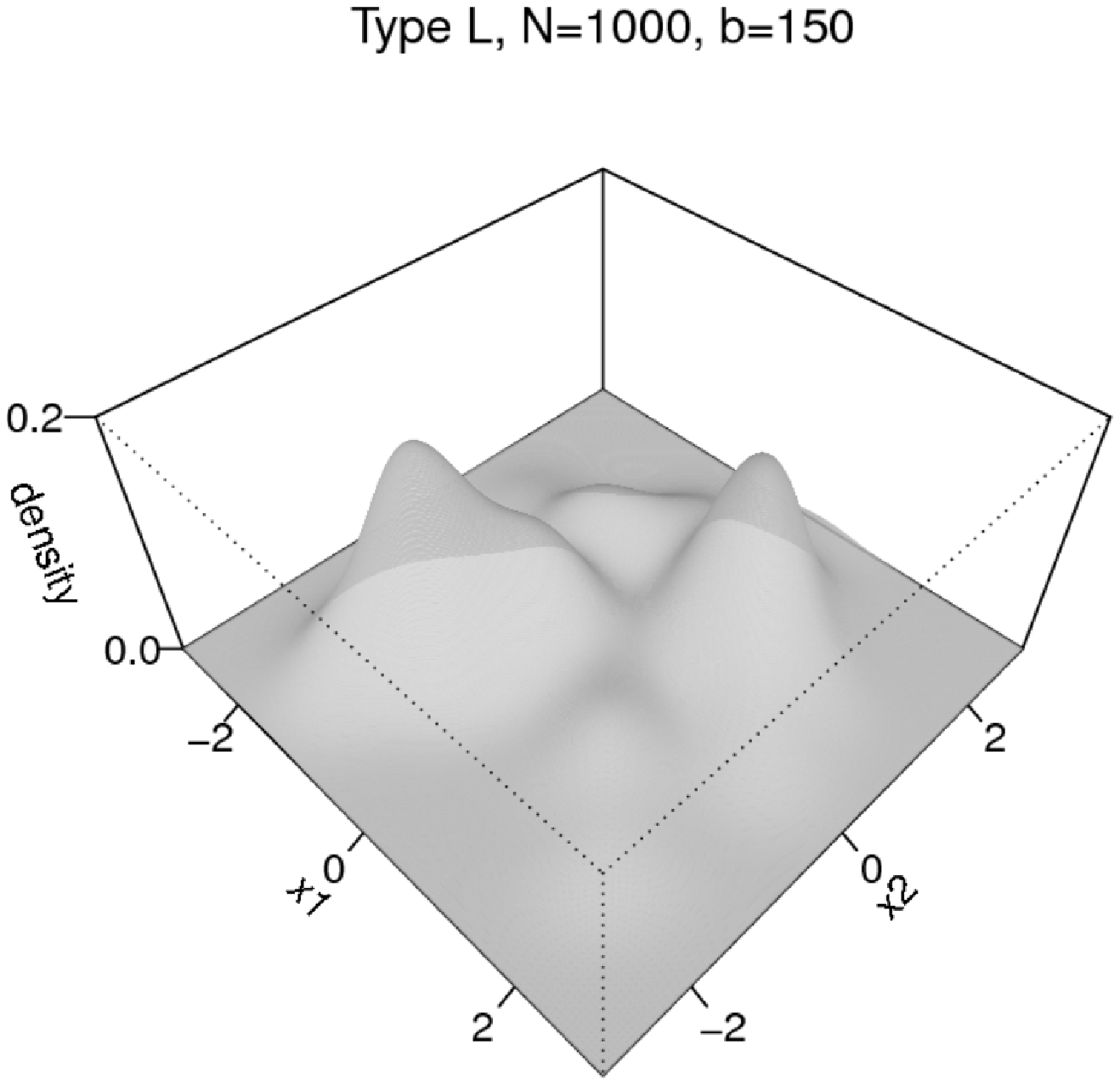}
        \mbox{(d)}
    \end{minipage}
\end{center}
\caption{[Simulation~1 (Type L, $N=1000$)] (a): Estimation error vs $b$ at $g=100$. (b): Estimation error vs $g$ at $b=150$. (c): Contour-plot with the original data points (red) and the data points choosen by the GA methods (blue). (d): Perspective-plot.} \label{sim1.N1000.L}
\end{figure}

Figure~\ref{sim1.N1000.L} summarizes the results in Table~\ref{L.results} for the case of $N=1000$ as an example, the counterpart of Figure~\ref{sim1.N1000.C}. From the panel (a) in Figure~\ref{sim1.N1000.L}, we observe that ISE$^{*}(100)$ takes the minimum value from $b=150$ to $b=300$. We assume that ISE$^{*}(100)$ would increase when $b$ exceeds 300 and approaches 1000. When $b$ is greater than the vicinity of $30$, our GA can outperform DPI estimator with its sample size $N=1000$ in terms of estimation error. The panel (b) in Figure~\ref{sim1.N1000.L} plots the value of ISE$^{*}(g)$ at every generation $g$ in the case of $b=150$. We observe that our proposed GA method of $b=150$ can outperform DPI estimator using the original sample of the size $N=1000$ in terms of estimation error when $g$ is greater than the vicinity of 20. The contour-plot in (c) of Figure~\ref{sim1.N1000.L} captures the shape of true density function and exhibits similar tendencies to the results in the case of Type~C. From the results in Table~\ref{L.results}, we observe our GA can outperform DPI estimator in Type~L only when $N=1000$, $b \ge 50$, and $g \ge 25$. It seems that Type~L is more difficult in making estimators using our GA than Type~C because Type~L requires a larger size of $b$ and $N$ than Type~C to achieve a smaller estimation error than DPI.

Table~\ref{tab.c.ratio.L} is the results in terms of the DCR for Type~L, being the counterpart of Table~\ref{tab.c.ratio.C}. We confirm the similar tendencies to Type~C and our GA can yield the smaller DCR than RSDE while simultaneously yielding smaller estimation errors.

We calculate MISE of our estimator in the same manner as Type~C. In the case of $(N, b) = (1000, 150)$, MISE$ \times 10^{5}$ (S.D. $ \times 10^{5}$) comes out to be $221$ (45), outperforming its competitors. The corresponding DCR (S.D.) is 0.0819 (0.0051).
\begin{table}
\begin{center}
{\scriptsize{
\begin{tabular}{lllllllllll}
\hline
\hline
$g$ & 1 & 25 & 50 & 75 & 100 & DPI & RSDE & DPI$^{*}$ & RSDE$^{*}$ \\
\hline
$\underline{N=200}$ & --- & --- & --- & --- & --- &  734\ (141) & 1299\ (524) & 613 & 1122 \\
$b=2$ & 2254 (399) & 1628 (315) & 1628 (315) & 1628 (315) & 1628 (315) \\

$b=5$ & 1672 (401) & 1372 (101) & 1378 (102) & 1378 (102) & 1378 (102) \\
$b=25$ & 1151 (240) & 974 (122) & 917 (129) & 887 (109) & 922 (81) \\
$b=50$ & 1179 (263) & 905 (114) & 880 (119) & 881 (91) & 846 (57) \\
$b=100$ & 957 (243) & 879 (130) & \underline{833} (86) & \underline{843} (81) & 871 (69) \\
$b=150$ & \underline{845} (180) & \underline{853} (59) & 867 (50) & 858 (71) & \underline{841} (81) \\
\hline
$\underline{N=400}$ & --- & --- & --- & --- & --- &  469\ (73) & 947 (284) & 406 & 834 \\
$b=2$ & 1801 (250) & 1443 (71) & 1443 (71) & 1443 (71) & 1443 (71) \\
$b=5$ & 1493 (297) & 1470 (368) & 1456 (357) & 1456 (357)  & 1456 (357) \\
$b=25$ & 1074 (195) & 737 (101) & 674 (117) & 639 (123) & 624 (101) \\
$b=50$ & 798 (206) & 636 (79) & 584 (75) & 598 (79) & 566 (54) \\
$b=100$ & 721 (149) & 634 (79) & 584 (77) & 562 (65) & 555 (38) \\
$b=150$ & \underline{615} (130) & \underline{578} (61) & \underline{553} (60) & \underline{562} (53) & \underline{527} (40) \\
\hline
$\underline{N=1000}$ & --- & --- & --- & --- & --- &  270\ (40) & 508 (103) & 223 & 455 \\
$b=2$ & 1980 (342) & 1486 (175) & 1486 (175) & 1486 (175) & 1486 (175) \\
$b=5$ & 1483 (213) & 1065 (88) & 1052 (79) & 1052 (79) & 1052 (79) \\
$b=25$ & 845 (115) & 433 (80) & 328 (63) & 325 (46) & 307 (37) \\
$b=50$ & 721 (80) & 294 (45) & 247 (31) & 224 (36) & 238 (43) \\
$b=100$ & 531 (84) & 263 (56) & 226 (31) & 222 (32) & 206 (26) \\
$b=150$ & \underline{470} (81) & \underline{237} (38) & \underline{196} (46) & \underline{187} (20) & \underline{179} (13)
 \\
\hline
\hline
\end{tabular}}}
\caption[] {[Simulation~1 (Type L)] Results of estimation error ISE$^{*}(g)$ $\times 10^{5}$ (S.D. $\times 10^{5}$). The numbers in the columns of DPI and RSDE are MISE $\times 10^{5}$ (S.D. $\times 10^{5}$). The numbers in the columns of DPI$^{*}$ and RSDE$^{*}$ are ISE $\times 10^{5}$ of each estimator calculated by the identical sample used in calculating ISE$^{*}$. The minimum values of ISE$^{*}(g)$ at each $g$ over the sizes of $b$ are underlined.} \label{L.results}
\end{center}
\end{table}
\begin{table}
\begin{center}
{\scriptsize{
\begin{tabular}{l|llll|l}
\hline
\hline
& (I) & (II) & (II)$/b\ \ \ $ & DCR.GA & DCR.RSDE \\
\hline
$\underline{N=200}$ & ---  & ---  & --- & --- & .2695\ (.0326) \\
$b=25$ & 14.60 (1.07) & 3.60 (0.84) & 0.14 & .0730 (.0053) & ---
 \\
$b=50$ & 25.20 (2.82) & 5.20 (0.91) & 0.10 & .1260 (.0141) & --- \\
$b=100$ & 36.70	(6.09) & 8.80 (2.14) & 	0.08 & .1835 (.0304) & --- \\
$b=150$ & 49.30	(3.83) & 11.90 (3.44) & 0.07 & .2465 (.0191) & --- \\
\hline
$\underline{N=400}$ & ---  & ---  & --- & --- & .2125\ (.0134) \\
$b=25$ & 17.00 (1.94) & 3.50 (1.08) & 	0.14 & .0425 (.0048) & --- \\
$b=50$ & 31.50 (2.36) & 4.50 (0.97) & 	0.09 & .0787 (.0059) & --- \\
$b=100$ & 51.90 (5.54) & 6.30 (1.63) & 	0.06 & .1297 (.0138) & --- \\
$b=150$ & 68.70 (5.16) & 8.10 (1.72) & 	0.05 & .1717 (.0129) & --- \\
\hline
$\underline{N=1000}$ & ---  & ---  & --- & --- & .1474\ (.0103) \\
$b=25$ & 17.50 (1.95) & 2.90 (0.56) & 	0.11 & .0175 (.0019) & --- \\
$b=50$ & 33.60 (2.41) & 4.10 (1.28) & 	0.08 & .0336 (.0024) & --- \\
$b=100$ & 59.30 (5.39) & 4.80 (0.78) & 	0.04 & .0593 (.0053) & --- \\
$b=150$ & 79.40 (7.64) & 6.50 (1.43) & 	0.04 & .0794 (.0076) & --- \\
\hline
\hline
\end{tabular}}}
\caption[]
{[Simulation~1 (Type L)] (I): The number of distinct data points used for estimating the density on the average (S.D.). (II): The maximum multiplicity of data points on the average (S.D.).} \label{tab.c.ratio.L}
\end{center}
\end{table}
\clearpage
\subsection{Simulation~2 (trivariate)} \label{Sim2}

In Simulation~2, we employ the following trivariate simulation setting for the true density, where the notation $N(\mu_{1}, \mu_{2}, \mu_{3}, \sigma_{1}^{2}, \sigma_{2}^{2}, \sigma_{1}^{3}, \rho_{12}, \rho_{13}, \rho_{23})$ represents the trivariate normal density with means $\mu_{1}$, $\mu_{2}$, and $\mu_{3}$, variances $\sigma_{1}^{2}$, $\sigma_{2}^{2}$, and $\sigma_{3}^{2}$, and correlation coefficients $\rho_{12}$, $\rho_{13}$, and $\rho_{23}$.
\\\\{\bf{Type $C$ trivariate. Trivariate skewed:}}\\
$\frac{1}{5} N\bigl(0, 0, 0, 1, 1, 1, 0, 0, 0 \bigr) + \frac{1}{5} N\bigl(\frac{1}{2}, \frac{1}{2}, \frac{1}{2}, \bigl(\frac{2}{3}\bigr)^{2}, \bigl(\frac{2}{3}\bigr)^{2}, \bigl(\frac{2}{3}\bigr)^{2}, 0, 0, 0 \bigr) + \frac{3}{5} N \bigl(\frac{13}{12}, \frac{13}{12}, \frac{13}{12}, \bigl(\frac{5}{9}\bigr)^{2}, \bigl(\frac{5}{9}\bigr)^{2}, \bigl(\frac{5}{9}\bigr)^{2}, 0, 0, 0 \bigr)$\\

\begin{figure}[]
\begin{center}
    \begin{minipage}[t]{0.49\hsize}
        \center
        \captionsetup{width=.95\linewidth}
        \includegraphics[width=\textwidth]{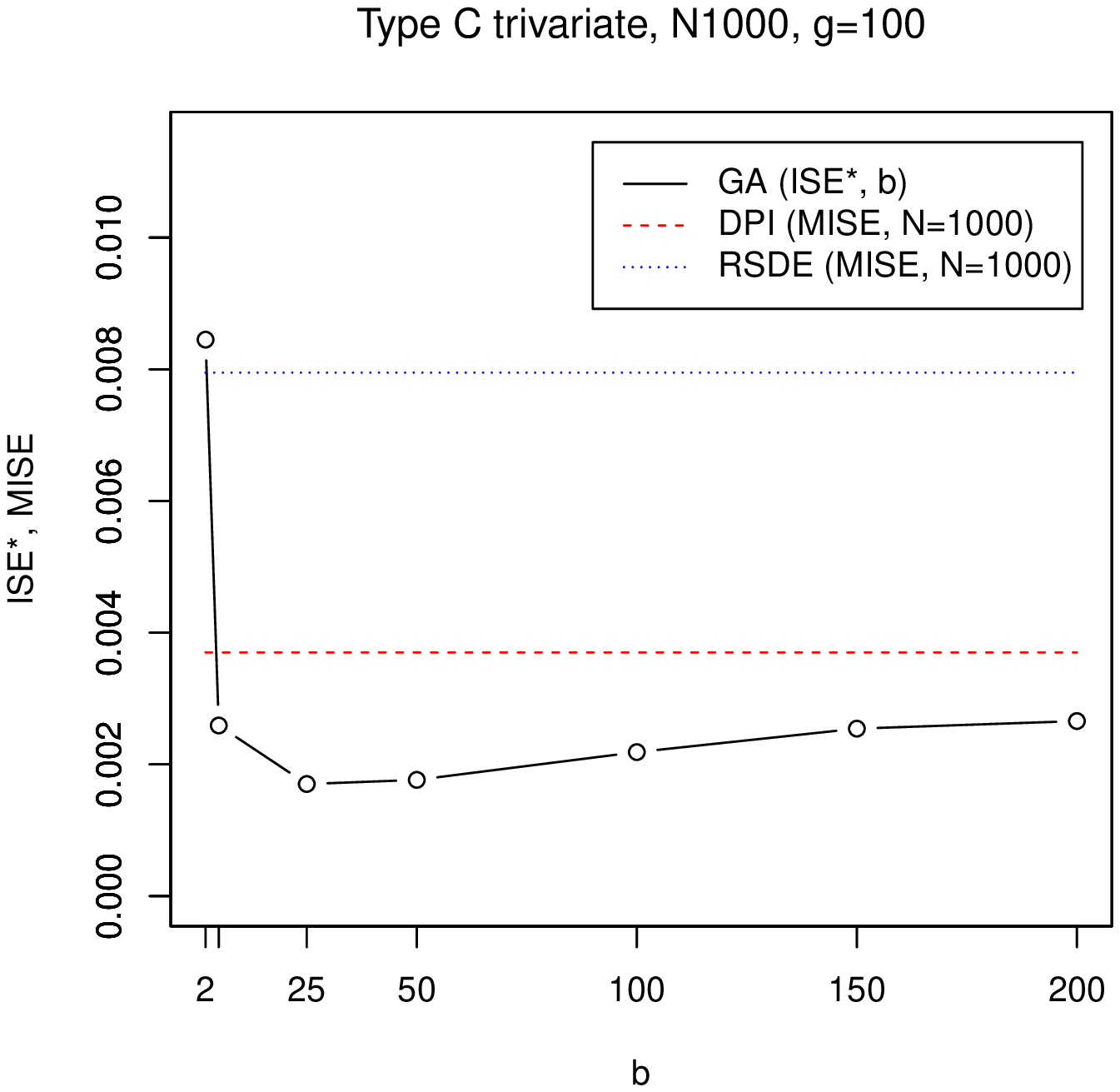}
        \mbox{(a)}
    \end{minipage}
   \begin{minipage}[t]{0.49\hsize}
        \center
        \captionsetup{width=.95\linewidth}
       \includegraphics[width=\textwidth]{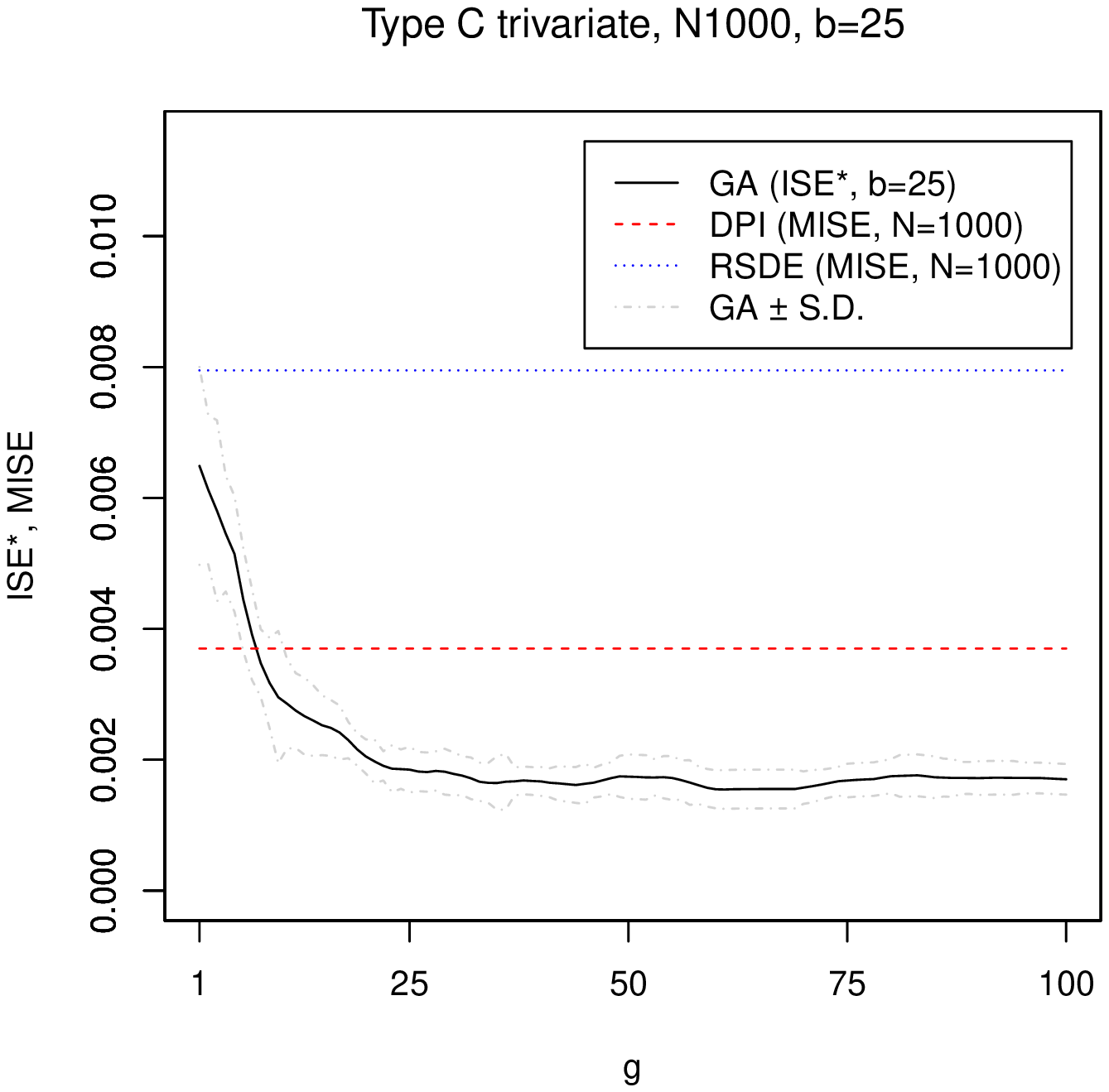}
        \mbox{(b)}
    \end{minipage}
\end{center}
\caption{[Simulation~2 (Type~C trivariate, $N=1000$)] (a): Estimation error vs $b$ at $g=100$. (b): Estimation error vs $g$ at $b=25$.} \label{sim2.N1000.C.3d}
\end{figure}

We implement the simulation in the same manner as Simulation~1 employing the same GA parameter settings $(B, G, p_{u}, p_{m}, p_{e})$. The numerical results are presented in Table~\ref{results.C.3d}, which are summarized in Figure~\ref{sim2.N1000.C.3d} for the case of $N=1000$. From the panel (a) in Figure~\ref{sim2.N1000.C.3d}, we observe that ISE$^{*}(100)$ takes the minimum value in the vicinity of $b=25$. When $b$ is greater than the vicinity of $5$, our GA can outperform the DPI estimator with its sample size $N=1000$ in terms of estimation error. The panel (b) in Figure~\ref{sim2.N1000.C.3d} plots the value of ISE$^{*}(g)$ at every generation $g$ in the case of $b=25$. We observe that our proposed GA method of $b=25$ can outperform the DPI estimator using the sample of the size $N=1000$ in terms of estimation error when $g$ is greater than the vicinity of 8. 

Comparing the results of ISE$^{*}(g)$ in Tables~\ref{C.results} and \ref{results.C.3d}, the speed of convergence with respect to $g$ reduces as the dimension increases, except for the cases of $(N, b) = (200, 2), (400, 2)$ and $(1000, 2)$. Looking at the numeric results in the cases of $N=200$ and $400$ in Table~\ref{results.C.3d}, we also observe that ISE$^{*}(100)$ is minimized at $b=5$ while in the case of Type~C bivariate, it is minimized at $b=25$ for the same size of $N$. We conjecture that a smaller number of training data points is sufficient to minimize estimation error in the larger dimension.

\begin{figure}[h]
\begin{center}
    \begin{minipage}[t]{0.49\hsize}
        \center
        \captionsetup{width=.95\linewidth}
        \includegraphics[width=\textwidth]{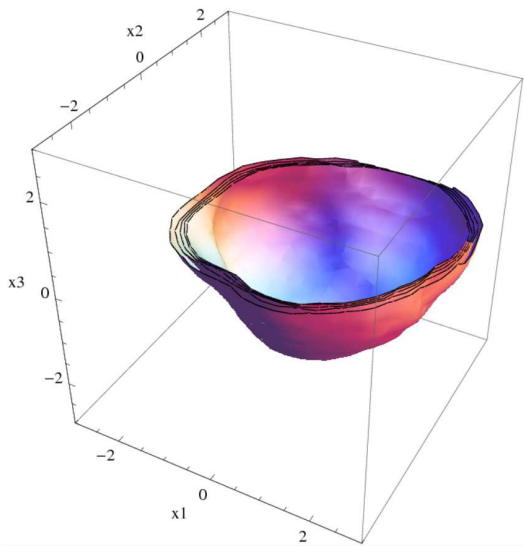}
        \mbox{(a)}
    \end{minipage}
    \begin{minipage}[t]{0.49\hsize}
        \center
        \captionsetup{width=.95\linewidth}
        \includegraphics[width=\textwidth]{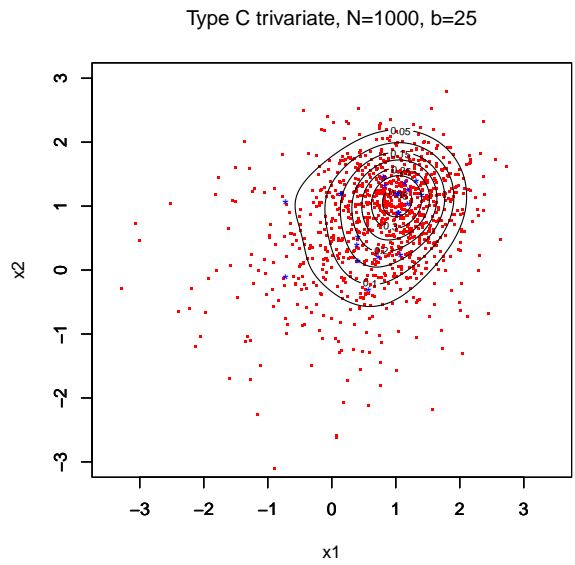}
        \mbox{(b)}
    \end{minipage}\\
    \begin{minipage}[t]{0.49\hsize}
        \center
        \captionsetup{width=.95\linewidth}
       \includegraphics[width=\textwidth]{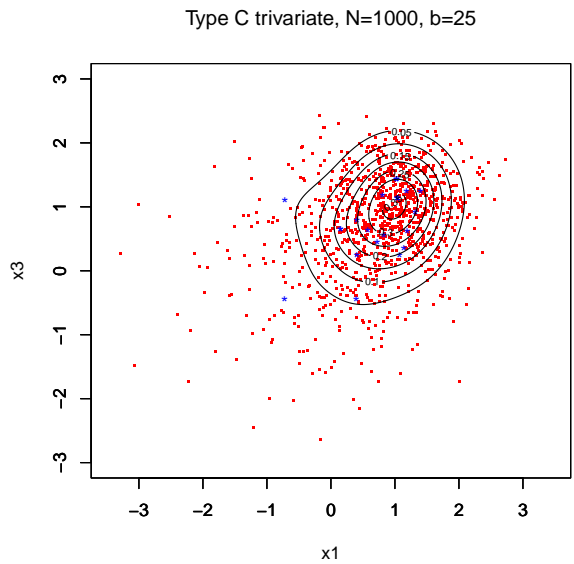}
       \mbox{(c)}
    \end{minipage}
    \begin{minipage}[t]{0.49\hsize}
        \center
        \captionsetup{width=.95\linewidth}
        \includegraphics[width=\textwidth]{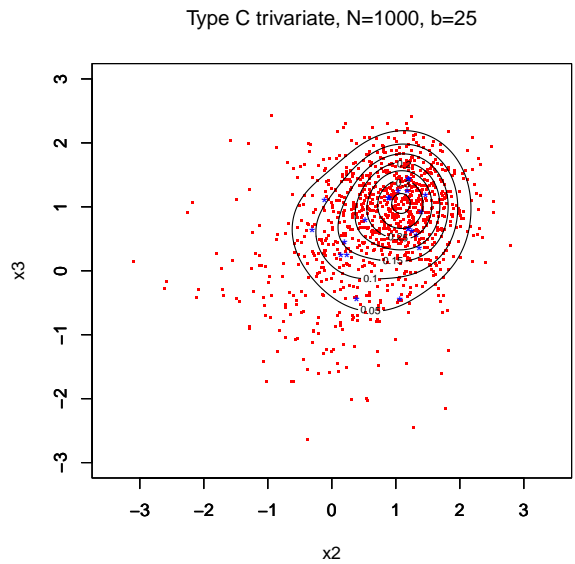}
        \mbox{(d)}
    \end{minipage}
\end{center}
\caption{[Simulation~2 (Type C, trivariate)] (a): 3-D contour-plot. (b): 2-D contour-plot on $x_{1}$ and $x_{2}$. (c): 2-D contour-plot on $x_{1}$ and $x_{3}$. (d): 2-D contour-plot on $x_{2}$ and $x_{3}$.} \label{sim2.N1000.C.3d.cnt}
\end{figure}

The panel (a) in Figure~\ref{sim2.N1000.C.3d.cnt} is the 3-D contour-plot of the result at the five density levels $0.1$, $0.2$, $0.3$, $0.4$ and $0.5$, which we draw using the bandwidth matrix $h^{2} \mathbf{I}_{3}$, $h=0.4910$, calculated by our GA and slice at $x_{3}=1.0$. The panels (b) (c) and (d) in the same Figure are the representations of the trivariate results by the 2-D contour-plots of $x_{1}$ vs $x_{2}$, $x_{1}$ vs $x_{3}$, and $x_{2}$ vs $x_{3}$ respectively, which we draw using the same bandwidth as that of (a). From the shape of the 2-D contour-plots, the three 2-D marginal densities of Type C trivariate are the same shape by symmetry, and the estimation by our GA appears to be working to some degree.

Table~\ref{tab.c.ratio.C.3d} presents the results in terms of the DCR for Type~C trivariate, being the counterpart of Tables~\ref{tab.c.ratio.C} and \ref{tab.c.ratio.L}. We confirm the similar tendencies to Type~C and L in terms of DCR's and our GA can yield the smaller DCR while simultaneously yielding smaller estimation error than RSDE. Observing the results in Table~\ref{tab.c.ratio.C} and \ref{tab.c.ratio.C.3d}, Type~C trivariate yields the larger DCR compared to Type~C when $N=200$ and $400$ for each size of $b$ while both are comparable in the case of $N=1000$.

We calculate MISE of our estimator in the same manner as simulation~1. In the case of $(N, b) = (1000, 25)$, MISE$ \times 10^{5}$ (S.D. $ \times 10^{5}$) comes out to be $162$ (71), outperforming its competitors. The corresponding DCR (S.D.) is 0.0157 (0.0020).
\begin{table}[h]
\begin{center}
{\scriptsize{
\begin{tabular}{llllllllll}
\hline
\hline
$g$ & 1 & 25 & 50 & 75 & 100 & DPI &  RSDE & DPI$^{*}$ & RSDE$^{*}$ \\
\hline
$\underline{N=200}$ & --- & --- & --- & --- & --- &  1053\ (252) & 1408\ (812) & 1039 & 2031 \\
$b=2$ & 1341 (263) & 795 (147)  &  795(147)  &  795(147) & 795 (147) \\

$b=5$ & 975 (382) & \underline{427} (63)  &  \underline{421} (53)  &  \underline{421} (53)  &  \underline{421} (53) \\
$b=25$ & 815 (186)  &  440 (158)  &  482 (159)  &  561 (150)  &  567 (145) \\
$b=50$ & \underline{604} (159)  &  524 (87)  &  726 (151)  &  846 (111)  &  842 (157) \\
$b=100$ & 678 (199)  & 738 (179) &  1137 (243)  &  1312 (204)  &  1391 (196) \\
$b=150$ & 741 (231)  &  895 (161)  &  1414 (139)  &  1446 (109)	&  1632 (114) \\
\hline
$\underline{N=400}$ & --- & --- & --- & --- & --- &  665\ (58) & 1041\ (248) & 704 & 1267 \\
$b=2$ & 1078 (331)  &  715 (93)  &  715 (93)  &  715 (93)  & 715 (93) \\

$b=5$ & 1067 (407)  &  \underline{369} (74)  &  \underline{362} (70)  &  \underline{362} (70)  &  \underline{362} (70) \\
$b=25$ & 753 (134)  & 409 (34)  &  419 (55)  &  438 (66)  &  464 (61) \\
$b=50$ & 648 (101)  &  411 (83)  &  470 (78)  & 517 (100)  &  511 (77) \\
$b=100$ & 682 (151)  &  493 (62)  &  555 (91)  &  554 (67)  &  638 (37) \\
$b=150$ & \underline{520} (79)  &  508 (58)  &  600 (66)  &  630 (59)  &  681 (68) \\
\hline
$\underline{N=1000}$ & --- & --- & --- & --- & --- &  369\ (55) & 794\ (177) & 361 & 950 \\
$b=2$ & 1076 (229)  &  845 (124)  & 845 (124)  &  845 (124)  &  845 (124) \\

$b=5$ & 846 (333)  &  251 (35)  &  259 (41)  &  259 (41)  &  259 (41) \\
$b=25$ & 648 (151)  &  184 (34)  &  \underline{174} (33)  & 168 (25)  & \underline{170} (23) \\
$b=50$ & 535 (119)  &  \underline{177} (38)  &  177 (48)  & \underline{167} (43) & 176 (32) \\
$b=100$ & 423 (104) & 193 (60)  & 209 (36)  &  217 (27)  & 218 (24) \\
$b=150$ & \underline{382} (94) & 201 (46)  & 220 (31)  &  243 (31)  &  254 (36) \\
\hline
\hline
\end{tabular}}}
\caption[]
{[Simulation~2 (Type C trivariate)] Results of estimation error ISE$^{*}(g)$ $\times 10^{5}$ (S.D. $\times 10^{5}$). The numbers in the columns of DPI and RSDE are MISE $\times 10^{5}$ (S.D. $\times 10^{5}$). The numbers in the columns of DPI$^{*}$ and RSDE$^{*}$ are ISE $\times 10^{5}$ of each estimator calculated by the identical sample used in calculating ISE$^{*}$. The minimum values of ISE$^{*}(g)$ at each $g$ over the sizes of $b$ are underlined.} \label{results.C.3d}
\end{center}
\end{table}
\clearpage
\begin{table}[h]
\begin{center}
{\scriptsize{
\begin{tabular}{l|llll|l}
\hline
\hline
& (I) & (II) & (II)$/b\ \ \ $ & DCR.GA & DCR.RSDE \\
\hline
$\underline{N=200}$ & ---  & ---  & --- & --- & .3775
\ (.0487) \\
$b=25$ & 14.40 (2.41) & 4.70 (0.82) & 0.18 & .0720 (.0120) & ---
 \\
$b=50$ & 26.70 (3.05) & 6.00 (1.88) & 0.12 & .1335 (.0152) & --- \\
$b=100$ & 43.40 (3.59) & 8.00 (2.10) & 0.08 & .2170 (.0179) & --- \\
$b=150$ & 55.50 (4.88) & 11.10 (1.72) & 0.07 & .2775 (.0244) & --- \\
\hline
$\underline{N=400}$ & ---  & ---  & --- & --- & .3135\ (.0298) \\
$b=25$ & 14.50 (1.90) & 4.50 (1.17) & 0.18 & .0362 (.0047) & --- \\
$b=50$ & 27.80 (3.42) & 6.00 (1.24) & 0.12 & .0695 (.0085) & --- \\
$b=100$ & 49.00 (4.32) & 8.00 (2.35) & 0.08 & .1225 (.0108) & --- \\
$b=150$ & 64.80 (6.62) & 11.50 (2.67) & 0.07 & .1620 (.0165) & --- \\
\hline
$\underline{N=1000}$ & ---  & ---  & --- & --- & .1857\ (.0111) \\
$b=25$ & 15.70 (1.82) & 5.30 (1.63) & 0.21 & .0157 (.0018) & --- \\
$b=50$ & 30.90 (2.99) & 5.60 (2.50) & 0.11 & .0309 (.0029) & --- \\
$b=100$ & 53.60 (5.77) & 7.10 (1.72) & 0.07 & .0536 (.0057) & --- \\
$b=150$ & 73.80 (5.49) & 9.00 (2.21) & 0.06 & .0738 (.0054) & --- \\
\hline
\hline
\end{tabular}}}
\caption[]
{[Simulation~2 (Type~C trivariate)] (I): The number of distinct data points used for estimating the density on the average (S.D.). (II): The maximum multiplicity of data points on the average (S.D.).} \label{tab.c.ratio.C.3d}
\end{center}
\end{table}
\clearpage
\subsection{Simulation~3 (Real data application)} \label{RealData}

We show a real data application of trivariate density estimation for our GA. We use {\it{abalone data set}}, originally from Nash et al. (1994), available in the UCI Machine Learning Repository. The dataset consists of the physical measurements of abalones collected in Tasmania, with each abalone being measured on eight attributes: Sex (male, female, infants), Length (mm), Diameter (mm), Height (mm), Whole weight (grams), Shucked weight (grams), Viscera weight (grams), Shell weight (grams), and Rings (integer). We construct the dataset consist of Diameter, Viscera weight and Shucked weight out of the eight attributes. The sample size $N$ is $1528$ for male abalones. In the estimation, we set $(b, B, G, p_{u}, p_{m}, p_{e})=(100, 50, 500, 0.475, 0.05, 0.1)$. We implement our GA once for the trivariate abalone dataset and obtain $\mathbf{H}^{*} = h^{2} \mathbf{I}_{3}$, $h = 0.025$ at the completion of our GA. In terms of data condensation, we obtain the following results; the number of different data points, 44; the maximum multiplicity, 6; and the DCR, 0.0288. For reference, RSDE brings DCR=0.1865.

\begin{figure}[h]
\begin{center}
    \begin{minipage}[t]{0.49\hsize}
        \center
        \captionsetup{width=.95\linewidth}
        \includegraphics[width=\textwidth]{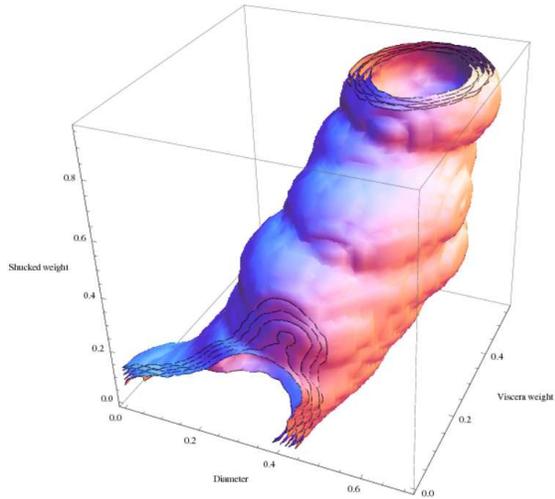}
        \mbox{(a)}
    \end{minipage}
    \begin{minipage}[t]{0.49\hsize}
        \center
        \captionsetup{width=.95\linewidth}
        \includegraphics[width=\textwidth]{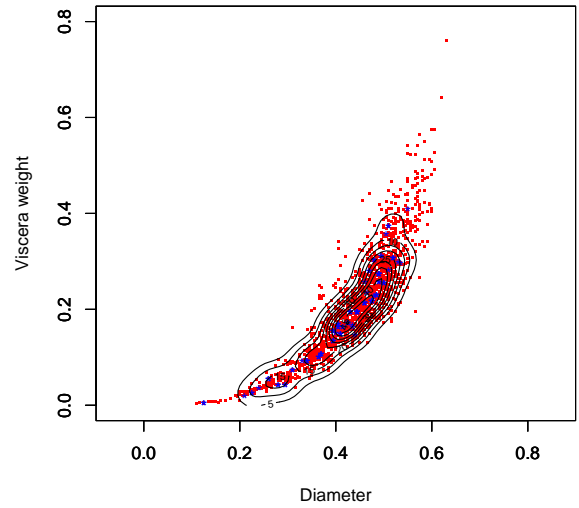}
        \mbox{(b)}
    \end{minipage}\\
    \begin{minipage}[t]{0.49\hsize}
        \center
        \captionsetup{width=.95\linewidth}
        \includegraphics[width=\textwidth]{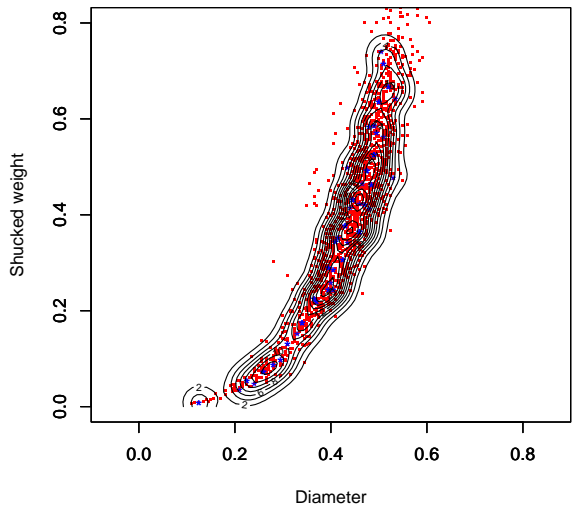}
        \mbox{(c)}
   \end{minipage}
    \begin{minipage}[t]{0.49\hsize}
        \center
        \captionsetup{width=.95\linewidth}
       \includegraphics[width=\textwidth]{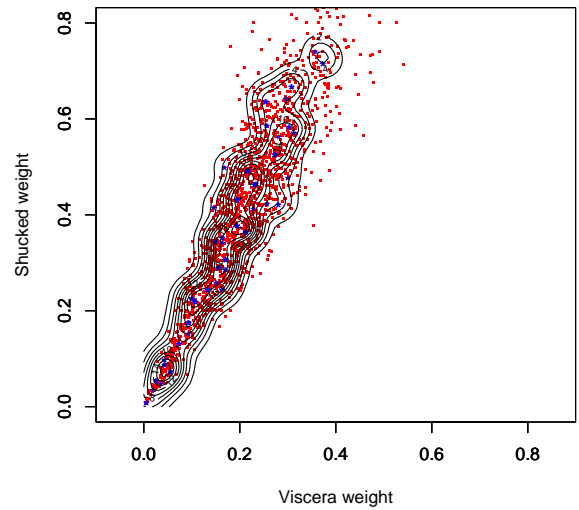}
       \mbox{(d)}
    \end{minipage}
\end{center}
\caption{[Simulation~3 (Real data application)] (a): 3-D contour-plot. (b): 2-D contour-plot on Dianeter and Viscera weight. (c): 2-D contour-plot on Diameter and Shucked weight. (d): 2-D contour-plot on Viscera weight and Shucked weight.} \label{abalone.3d}
\end{figure}

The panel (a) in Figure~\ref{abalone.3d} is the 3-D contour-plot of the result at the five density levels $0.00001$, $0.0001$, $0.001$, $0.01$ and $0.1$. The panels (b) (c) and (d) in the same Figure are the representations of the trivariate results by the 2-D contour-plots of Diameter vs Viscera weight, Diameter vs Shucked weight, and Viscera weight vs Shucked weight respectively, which we draw by using the same bandwidth matrix as that of (a). The red points in the 2-D contour-plots designate the original data points, while the blue ones are chosen by our GA for the estimation. From the shape of the 2-D contour-plots, the estimation by our GA appears to be working. From the 2-D contour-plots (b), (c) and (d) tell us that our GA generally chooses data points along with the mountain ridges of the contour-plots. This tendency is also observed in RSDE (Girolami and He 2003, p.1256) and SMA (Nishida and Naito 2021).

\begin{figure}[h]
\begin{center}
    \begin{minipage}[t]{0.49\hsize}
        \center
        \captionsetup{width=.95\linewidth}
        \includegraphics[width=\textwidth]{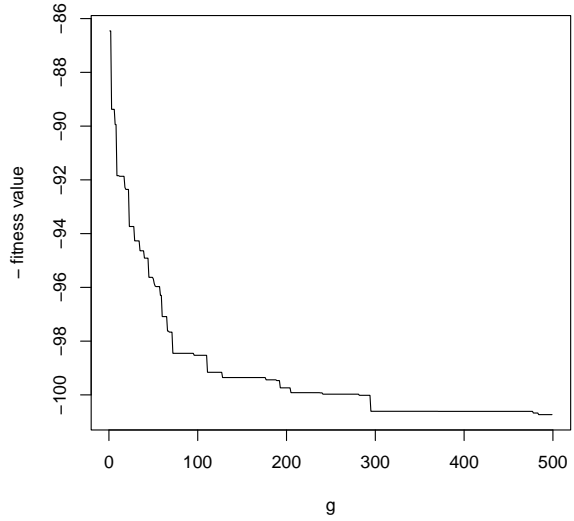}
        \mbox{(a)} 
    \end{minipage}
    \begin{minipage}[t]{0.49\hsize}
        \center
        \captionsetup{width=.95\linewidth}
        \includegraphics[width=\textwidth]{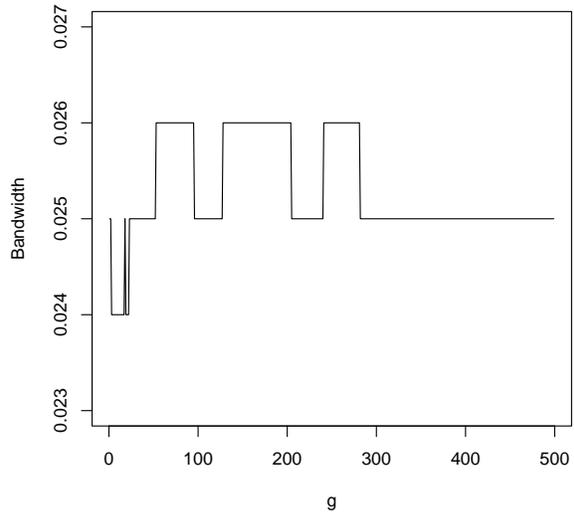}
        \mbox{(b)} 
    \end{minipage}
    \begin{minipage}[t]{0.49\hsize}
        \center
        \captionsetup{width=.95\linewidth}
        \includegraphics[width=\textwidth]{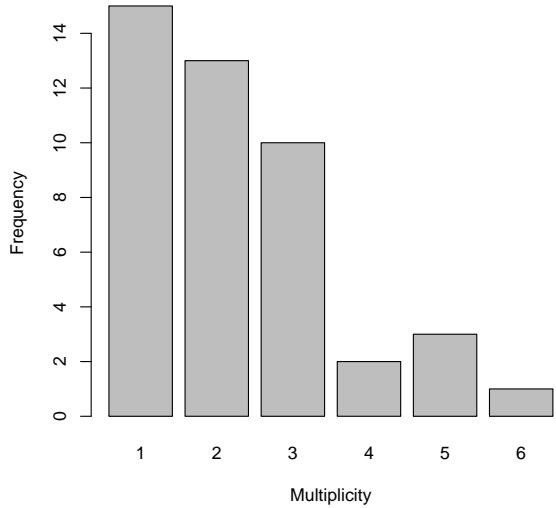}
        \mbox{(c)}
    \end{minipage}
    \begin{minipage}[t]{0.49\hsize}
        \center
        \captionsetup{width=.95\linewidth}
        \includegraphics[width=\textwidth]{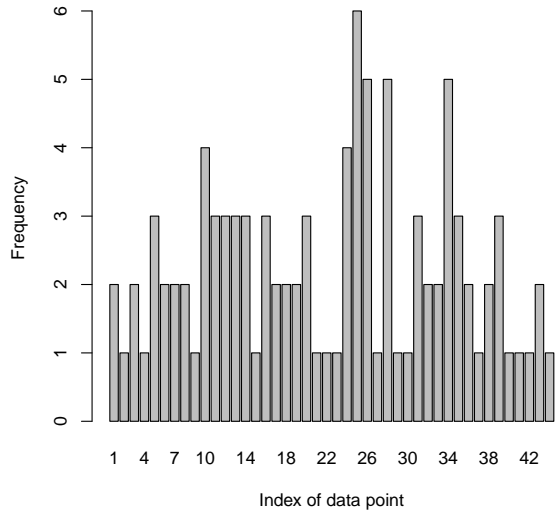}
        \mbox{(d)}
    \end{minipage}
\end{center}
\caption{[Results 2 (real data application)] (a): Plot of the (minus) fitness values at every $g$. (b): Plot of the bandwidths at every $g$. (c): Histgram of the multiplicity of the data points. (d): Frequency plots of the data points selected by our GA. The data points are indexed in the ascending order of distance from the origin.} \label{abalone.3d.supp}
\end{figure}

The panels (a) (b) (c) and (d) in Figure~\ref{abalone.3d.supp} are the plot of the (minus) fitness values in \eqref{def.CV} at each generation $g$, the plot of the bandwidths at each generation $g$, the histgram of the multiplicity of each data point, and the frequency plots of the data points selected by our GA respectively. In the panel (d), the data points are indexed in the ascending order of distance from the origin to illustrate the spatial distribution of weights. The panel (a) tells us our GA reduces the (minus) fitness value as $g$ progresses. The panel (b) tells us that the bandwidth is no longer updated when $g$ is greater than 300. The panel (c) tells us that the only data point (Diameter, Viscera weight, Shucked weight) $ = (0.445, 0.1945, 0.4315)$ is chosen 6 times. The panel (d) tells us that greater weights are placed on those in the vicinity of the data point indexed as 25. It also tells us that the data points with unit weight are uniformly distributed over the indices.
\clearpage
\section{Discussion} \label{Dis}

In this study, we propose a data condensation method for KDE by GA. In the initial generation of our GA, we first construct multiple subsamples of a given size with replacement from the original sample, where each subsample and each of the constituting data point is called a chromosome and gene, respectively. In the subsequent generation, we evaluate the chromosomes in terms of fitness, and some chromosomes are inherited to the next generation by the elite selection rule. In line with elite selection, a pair of chromosomes, paired in order of the best fitness value, breed two new chromosomes by crossover, mutation, and reproduction and some of the new chromosomes, whose recalculate fitness values predominate those of the rest, are also inherited to the next generation in such a way that the total number of chromosomes inherited to the next remains unchanged over generation. This process is repeated generation by generation until the terminating condition is fulfilled and we finally obtain the KDE using the best subsample along with the best smoothing parameter.

We validate the performance of our GA by simulations and confirm that it can yield the KDE better than DPI and RSDE in terms of estimation error and DCR in many situations. In addition to the simulation studies, we also conduct the sensitivity analysis of the tuning parameters $B, p_{u}, p_{m}$, and $p_{e}$ to the performances of the resulting density estimator although we omit presenting the detailed results in this study for its brevity. We confirm that the sizes of $p_{m}$ and $p_{u}$ are influential to the results; if improperly selected, the speed of convergence becomes slow. Hence, we follow the practice adopted in the literature $p_{m} \le 0.05 \ll p_{u} < 0.5$ in this study (see Remark~\ref{remark.sensitivity}). We also confirm that the impact of the size of $p_{e} B$ is nonsignificant unless $p_{e} B$ is set to be miniscule. As for the size of $B$, if the size of $B$ is miniscule, the diversity of the chromosomes is reduced. We confirm that $B$ ranging from $20$ to $60$ is proper for the original sample sizes given in our simulation studies.

Some variants of our GA are possible, such as employing {\it{$k$-point crossover}} in the stage of crossover or performing {\it{roulette selection}}, {\it{tournament selection}}, and {\it{ranking selection}} in the stage of selection (e.g., Haupt and Haupt 2004; Sivanandam and Deepa 2008). Our supplemental simulation studies show that these variants are not better than our proposed GA method in terms of the speed of convergence; hence, we do not present the detailed results.

This study highlights three important issues associated with our GA. First, it is noteworthy that our GA can be superior to DPI in terms of estimation error even though our method employs a scalar bandwidth matrix, that is, the simpler matrix form. Second, our simulation results tell us how we determine the size of $b$, which is equivalent to how we choose a combination of the data points of the size $b$ from the original sample, that plays a role of smoothing parameter as well as the well-known smoothing parameters such as bandwidth and weighting parameter. Our study shows that the optimal $b$ is determined by the balance of effects between the training and test data points in their numbers. Third, our GA is similar to RSDE in that both the methods ultimately aim to find the optimal weighting parameters assigned to each data point (see Remark~\ref{remark.weight}). In the case of RSDE, it is required to compute the initial bandwidth matrix $\mathbf{H}_{R}$, which is calculated in this study by the least squared cross-validation method under the condition $\alpha_{1} = \alpha_{2}= \cdots = \alpha_{N} = 1/N$ following Girolami and He (2003). This initial bandwidth matrix does not necessarily generate the best performance of estimation error for RSDE among the possible combinations of $(\alpha_{1}, \alpha_{2}, ..., \alpha_{N}, \mathbf{H}_{R})$ because the optimization of $\mathbf{H}_{R}$ in the initial stage does not anticipate optimizing $\alpha_{1}, \alpha_{2}, ..., \alpha_{N}$. Better initial bandwidth matrices would probably exist, although they cannot be found. In contrast, our GA is different in that our method can update the optimal bandwidth at every stage. It also looks like our GA performs significantly better than RSDE in terms of DCR.

For further studies, we will work toward finding the optimal size of subsamples and apply our GA to kernel regression estimation.

\clearpage
\section*{Acknowledgements}
The author gratefully acknowledges the financial support from KAKENHI 19K11851.

\end{document}